\documentclass[preprint,times]{aastex61}
\setcitestyle{square,numbers}
\usepackage{amsmath} 
\usepackage{amssymb}
\usepackage[OT1]{fontenc}

\newcommand{\xmm}{\emph{XMM-Newton}}
\newcommand{\chan}{\emph{Chandra}}
\renewcommand{\thefootnote}{\Alph{footnote}}

\accepted{to \emph{Nature Astronomy}}

\shorttitle{Anisotropic winds in new colliding-wind binary}
\shortauthors{Callingham et al.} 

\begin{document}

\title{Anisotropic winds in Wolf-Rayet binary identify potential gamma-ray burst progenitor}

\correspondingauthor{J.~R.~Callingham}
\email{callingham@astron.nl}

\author[0000-0002-7167-1819]{J.~R.~Callingham}
\affil{ASTRON, Netherlands Institute for Radio Astronomy, Oude Hoogeveensedijk 4, 7991PD, Dwingeloo, The Netherlands}
\affiliation{Sydney Institute for Astronomy (SIfA), School of Physics, The University of Sydney, NSW 2006, Australia}

\author[0000-0001-7026-6291]{P.~G.~Tuthill}
\affiliation{Sydney Institute for Astronomy (SIfA), School of Physics, The University of Sydney, NSW 2006, Australia}

\author[0000-0003-2595-9114]{B.~J.~S.~Pope}
\affiliation{Sydney Institute for Astronomy (SIfA), School of Physics, The University of Sydney, NSW 2006, Australia}
\affiliation{Center for Cosmology and Particle Physics, Department of Physics, New York University, 726 Broadway, New York, NY 10003, USA}
\affiliation{NASA Sagan Fellow}

\author{P.~M.~Williams}
\affiliation{Institute for Astronomy, University of Edinburgh, Royal Observatory, Edinburgh EH9 3HJ, UK}

\author{P.~A.~Crowther}
\affiliation{Department of Physics \& Astronomy, University of Sheffield, Sheffield, S3 7RH, UK}

\author{M.~Edwards}
\affiliation{Sydney Institute for Astronomy (SIfA), School of Physics, The University of Sydney, NSW 2006, Australia}

\author{B.~Norris}
\affiliation{Sydney Institute for Astronomy (SIfA), School of Physics, The University of Sydney, NSW 2006, Australia}

\author{L.~Kedziora-Chudczer}
\affiliation{School of Physics, University of New South Wales, NSW 2052, Australia}

\begin{abstract}

\noindent\textbf{The massive evolved Wolf-Rayet stars sometimes occur in colliding-wind binary systems in which dust plumes are formed as a result of the collision of stellar winds \citep{2007ARA&A..45..177C}. These structures are known to encode the parameters of the binary orbit and winds \citep{1999Natur.398..487T,1999ApJ...525L..97M,tuthill2008}. Here, we report observations of a previously undiscovered Wolf-Rayet system, 2XMM~J160050.7--514245, with a spectroscopically determined wind speed of $\approx$\,3400~\,km\,s$^{-1}$. In the thermal infrared, the system is adorned with a prominent $\approx$\,12$''$~spiral dust plume, revealed by proper motion studies to be expanding at only $\approx\,$570\,km\,s$^{-1}$. As the dust and gas appear coeval, these observations are inconsistent with existing models of the dynamics of such colliding wind systems \citep{2002MNRAS.329..897H,2015MNRAS.450.2551M,2017ApJ...835L..31L}. We propose that this contradiction can be resolved if the system is capable of launching extremely anisotropic winds. Near-critical stellar rotation is known to drive such winds \citep{2004A&A...418..639A,deMink2013}, suggesting this Wolf-Rayet system as a potential Galactic progenitor system for long-duration gamma-ray bursts.}

\end{abstract}

\section*{Main Body} \label{sec:body}

Wolf-Rayet (WR) stars represent the final stage of the evolution of the most massive stars before ending their lives as supernovae. Late-type carbon-rich WR stars with binary companions have the potential to produce spiral ``Pinwheel'' patterns in which dust forms at the interface between the colliding stellar winds \citep{1999Natur.398..487T,1990MNRAS.243..662W}. As the orbital motion entangles the winds, the form of the plume encodes the primary wind and orbital parameters, forming rare and powerful laboratories for testing our understanding of the mass-loss in WR stars. For well studied Pinwheels such as WR 104 \citep{1999Natur.398..487T,tuthill2008}, WR 98a \citep{1999ApJ...525L..97M} and WR~140 \citep{Monnier2002,Williams2009}, nearly complete solutions can be obtained that tightly constrain the wind speeds, wind-momentum ratio, and orbital parameters. For WR~104 and WR~140, the dust (studied by its proper motion in the thermal infrared) and the gas (the dominant wind component in the line of sight revealed by spectroscopy) have been shown to be co-moving, as expected for spherical stellar winds. 

Additionally, WR stars play a significant role in the chemistry and kinetic energy budget of the interstellar medium \citep{2007ARA&A..45..177C}, and are considered to be likely progenitors to long-duration gamma-ray bursts (GRBs) \citep{2006ApJ...637..914W,2008A&A...484..831D}. A key ingredient in most models for the production of long-duration GRBs is rapid rotation of the WR progenitor star \citep{2006ApJ...637..914W}. For stars that have solar-like metallicity, as observed for most Galactic WR stars \citep{2005A&A...429..581M,2016A&A...588A..50M}, line-driven winds rapidly rob the star of angular momentum. One channel to produce near critical-rotation of the WR star before undergoing a core-collapse supernova is through binary interaction\citep{2005ApJ...623..302F}. Unfortunately debates over the role of rotation remain largely in the domain of theory as it has proven extremely difficult to place any observational constraints on the rotation of WR stars. Because WR spectra are generally formed in their extended dense winds \citep{2014A&A...562A.118S}, obtaining rotational velocity from fitting the rotational broadening of photospheric lines is generally not possible. Therefore, there is no unambiguous detection of a rapidly rotating WR star in the Milky Way.

\begin{figure}
\begin{center}
\includegraphics[scale=0.55]{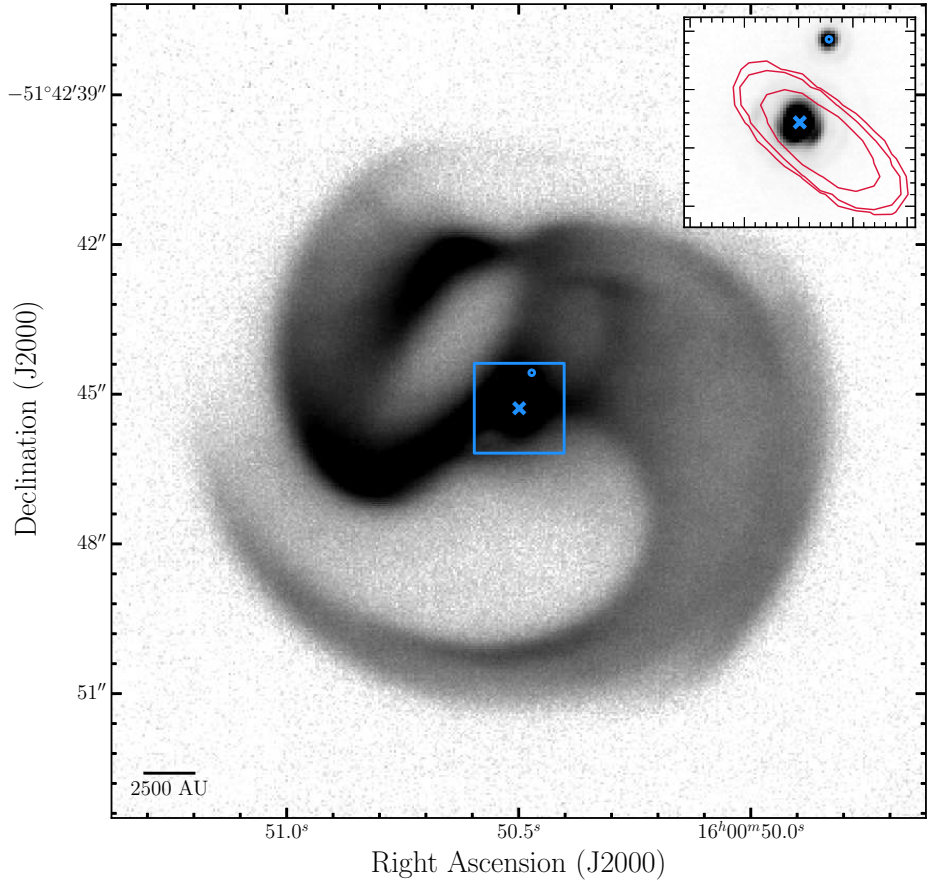}
 \caption{VISIR 8.9\,$\mu$m image of Apep taken on 2016 August 13, displaying the exotic dust pattern being sculpted by the system. The 2.24\,$\mu m$ NACO image of the region bounded by the blue box, of dimension $1.8'' \times 1.8''$, is shown in the upper right corner. The position of the Central Engine and the northern companion identified in the NACO image are indicated by the blue cross and circle, respectively. The over-plotted red contours are from our 19.7\,GHz ATCA observation, with contour levels correponding to 5, 10 and 50 times the rms noise (8\,$\times 10^{-2}$\,mJy\,beam$^{-1}$). The dimensions of the ATCA synthesised beam are $0.74'' \times 0.29''$, with a position angle of 50.1$^{\circ}$. The log stretch on the VISIR image is chosen to accentuate the dust pattern, ranging from 0.3 to 3\,mJy\,pixel$^{-1}$, with $\approx$\,20\% of the total $\approx$\,60\,Jy flux coming from the area bounded by the blue box. A scale bar of 2500 AU, for a distance of 2.4\,kpc, is provided at the bottom left corner of the plot.}
\label{fig:visir_img}
\end{center}
\end{figure}

2XMM~J160050.7-514245 (RA: 16:00:50.48, Dec: -51:42:45.4; J2000) was first noted as a high-luminosity outlier in our Galactic plane X-ray and radio survey, and revealed as a truly exceptional object upon considering its infrared spectral energy distribution (SED), where it brightens from an apparent magnitude of 6.4 at 2.2\,$\mu$m \citep{Skrutskie2006} to $-$2.4 at 22\,$\mu$m\citep{Wright2010}, with both measurements on the Vega system. Such dramatic brightening to the far infrared indicates the presence of luminous objects embedded within an extremely dusty environment. To explore the morphology of the dust nebula, and so divine the nature of the source, we observed the object with the mid-infrared camera VISIR on the European Southern Observatory's (ESO) Very Large Telescope (VLT) on 2016 August 13. The spectacular dust plume revealed at 8.9\,$\mu$m is shown in Figure~\ref{fig:visir_img}, exhibiting a form strongly reminiscent of the Archimedean spirals produced by the WR Pinwheels. Perhaps the closest resemblance is with the more complex forms exhibited by WR~140 \citep{Monnier2002,Williams2009} rather than the prototype system WR 104 \citep{1999Natur.398..487T}. Since only cumbersome catalog names like 2XMM~J160050.7-514245 are available for the system, we here adopt the moniker ``Apep''\footnote{The serpent deity from Egyptian mythology; mortal enemy of the Sun god Ra. We think this is an apt allusion to the image which evokes a star embattled within a serpent's coils.} after the sinuous form of this infrared plume. 

To identify the stellar spectral types and search for the expected fast ($\gtrsim 1000$\,km\,s$^{-1}$) wind in the system, we observed Apep using the near-infrared NACO camera and integral field spectrometer SINFONI, also on the VLT. The 2.24\,$\mu m$ NACO observation (Figure~\ref{fig:visir_img}, inset) resolves Apep into a 0.739$''$\,$\pm$\,0.002$''$ binary with a fainter companion to the North. When registered against the VISIR images, the brighter southern component (hereafter ``the Central Engine'') was found to exactly coincide with the mid-infrared peak central to the major structural elements of the dust plume. The fainter northern companion is just visible as a minor asymmetric feature at the corresponding location just outside the mid-infrared core.

Several further lines of evidence imply that the northern companion is unlikely to play any significant role in sculpting the dust plume. Assuming a distance of 2.4\,kpc (see Supplementary Information section\,\ref{sec:dist}), the companion is $\approx$\,1700\,AU from the Central Engine: several orders of magnitude wider than the normal range for Pinwheel binaries \citep{tuthill2008}. The corresponding orbital period of at least $10^4$ years could not wrap winds into a spiral in the manner of a Pinwheel system (the required windspeeds would be $<<$~1\,km\,s$^{-1}$) \citep{1999ApJ...525L..97M,tuthill2008}. Furthermore, the non-thermal radio emission from our 19.7\,GHz Australia Telescope Compact Array (ATCA) observations of the system also falls entirely on the Central Engine, with no offset to the North.

\begin{figure*}
\begin{center}$
\begin{array}{c}
\includegraphics[scale=0.5]{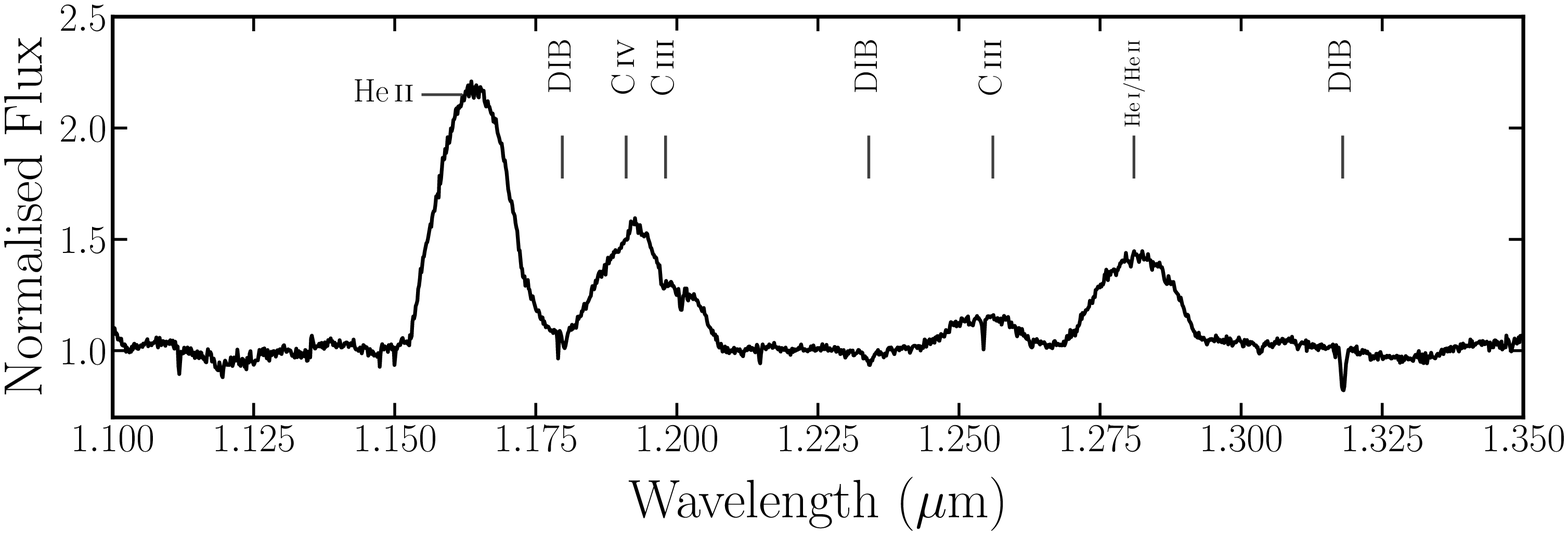} \\
\includegraphics[scale=0.5]{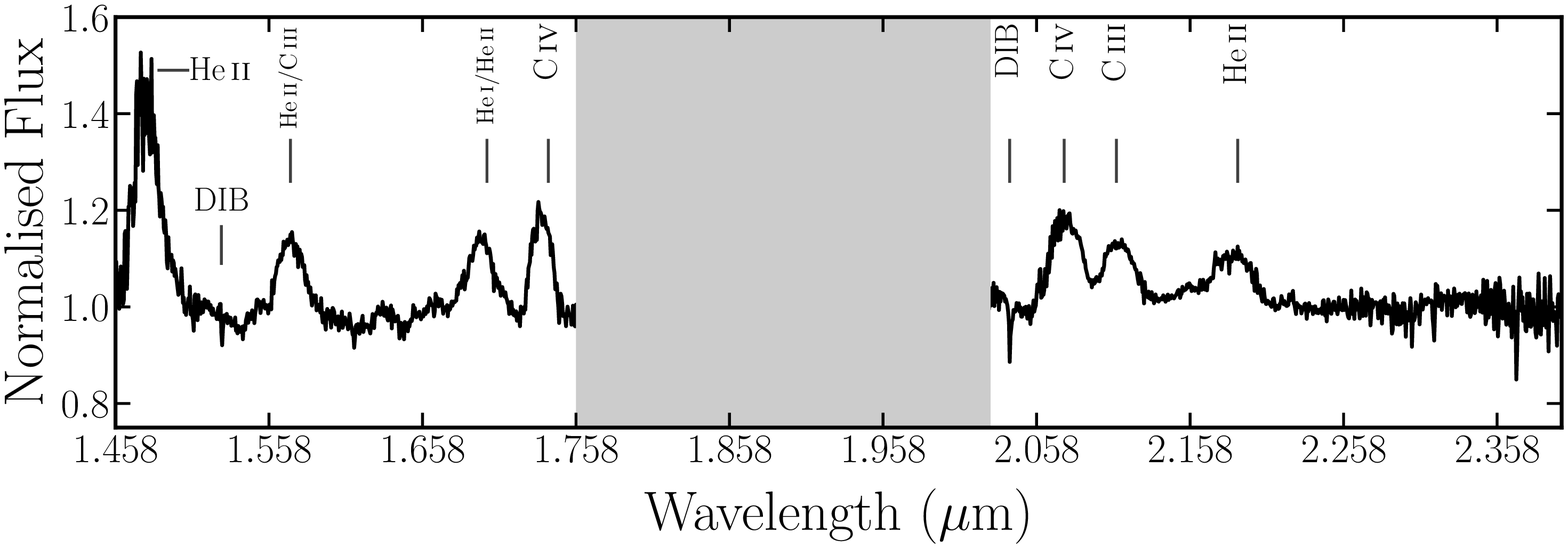}\\
\end{array}$
 \caption{SINFONI $J$-band (top) and $H$+$K$-band (bottom) spectra for the Central Engine at the centre of NACO image shown in the inset of Figure~\ref{fig:visir_img}. Prominent emission lines are labelled and the wavelength range where telluric correction was not possible between the $H$-~and $K$-bands is indicated in gray. Known and suggested diffuse interstellar absorption bands are labelled by `DIB', with the DIB line indicated at 2.02\,$\mu$m likely to be the first DIB detected in $K$-band.}
\label{fig:sinfoni_spec}
\end{center}
\end{figure*}

\begin{figure*}
\begin{center}
    \includegraphics[scale=0.5]{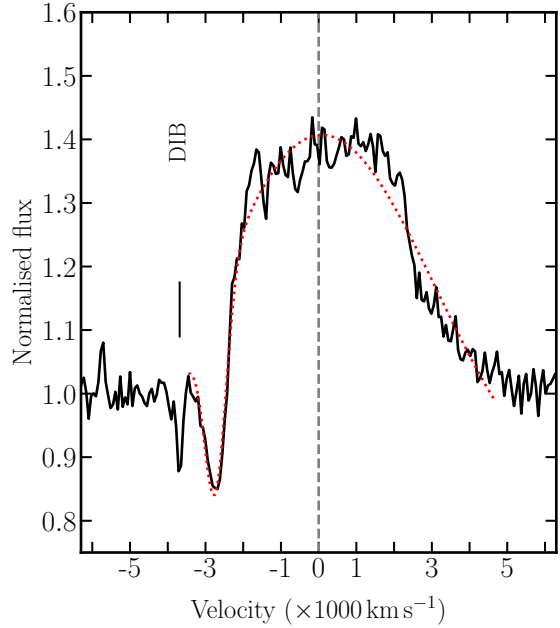}
    \caption{The 1.083\,$\mu$m\,He\,I line from the continuum-corrected IRIS2 long-slit $J_{s}$-band spectrum of Apep. The known diffuse interstellar band is identified by the label `DIB'. The fit to the P~Cygni profile is shown by the red-dashed curve, providing a measurement of a terminal velocity $v_{\infty} = 3400 \pm 200$\,km\,s$^{-1}$.}
    \label{fig:he1_line}
   \end{center}
\end{figure*}

The SINFONI instrument was able to isolate spectra for both the Central Engine and northern companion, with the $J$- and $H$+$K$-band spectra of the Central Engine presented in Figure~\ref{fig:sinfoni_spec}. The presence of a WR star in the Central Engine is confirmed by the absence of hydrogen lines and the characteristic broad helium and carbon lines \citep{1968MNRAS.138..109S}. The ratio of the C\,III and C\,IV lines, and the abnormal strength of the He\,II lines, suggests the presence of a carbon-rich WR star (WC) with a spectral type of WC7 or a WR star in the brief transitory phase between nitrogen and carbon-rich (WN/WC) \citep{Crowther2006,2018MNRAS.473.2853R}. Both subtypes have winds\citep{2007ARA&A..45..177C} $\gtrsim 1700$\,km\,s$^{-1}$. We spectroscopically measured the windspeed of the Central Engine of Apep via the 1.083\,$\mu$m\,He\,I line using the long-slit spectrograph IRIS2 on the 3.9\,m Anglo-Australian Telescope (AAT). As shown in Figure~\ref{fig:he1_line}, a fit to the P~Cygni profile of the 1.083\,$\mu$m\,He\,I line provides a direct spectroscopic measurement that a wind exists in the system with a terminal velocity $v_{\infty} =  3400 \pm 200$\,km\,s$^{-1}$.

Because fast WR winds usually result in proper motions that are readily apparent on short timescales, an additional VISIR epoch was observed on 1 August 2017, approximately one year after the original VISIR image. Proper motions revealed a pure radial expansion of the dust, confirming our identification of a Pinwheel-type plume in Apep. However, expectations for this displacement were confounded: structures were measured to be uniformly inflating at only 50\,$\pm$\,6\,mas\,yr$^{-1}$ (see Supplementary Information section\,\ref{sec:diff}). At $2.4$\,kpc this corresponds to an expansion velocity of 570\,$\pm$\,70\,km\,s$^{-1}$ —- nearly an order of magnitude slower than the fast wind measured spectroscopically, and well below lower bounds for any WR wind \citep{2007ARA&A..45..177C} for which limits can be traced to fundamental escape velocity arguments. The classical prototype Pinwheels exhibit no such discrepancy: angular and spectroscopic expansion speeds are consistent and simply related by the distance. Newly formed dust inherits the motion of the gas in which it formed, in this case the shock-compressed stellar wind. The latter depends on the densities and winds of the two stars local to the wind-collision region. In the case of isotropic stellar winds, this can be calculated from mass-loss rates and wind terminal velocities\citep{Canto1996}. Owing to mass loading, the velocity is less than that of the stellar winds, about 80 per cent in the case of WR 140, as has been observed from spectroscopy of `sub-peaks' on emission lines\citep{Marchenko2003}, but it is not typically much slower than this. Radiation pressure on the newly formed grains then accelerates them to velocities closer to that of the stellar winds\citep{Williams2009}.

Several straightforward arguments might attempt to explain this deep contradiction in the properties of the Apep wind. The measured dust and gas velocities would be reconciled if Apep were nearly ten times further away than indicated. However, this scenario is ruled out by kinematic distance limits (see Supplementary Information section\,\ref{sec:dist}), and furthermore it would boost intrinsic luminosities by a factor of 100 making Apep by far the brightest persistent X-ray and radio colliding wind-binary (CWB) system in the Galaxy \citep{2018MNRAS.474.3228P} (more luminous by an order of magnitude in radio emission than $\eta$~Carinae when it is not in outburst \citep{2013A&A...558A..28D}). Although it is difficult to conceive of an environment capable of mounting sustained resistance to a WR wind, the idea of some form of wind braking or pressure confinement might be proposed. Apart from the challenge of finding a mechanism to contain the momentum (by a factor of 10) of the most powerful stellar winds known, the clean form and detailed sharp structure in the dust plume of Figure~\ref{fig:visir_img} argue strongly against a confinement argument. The medium performing the braking would have to be highly uniform or else it would disrupt the plume's symmetry, and it would likely be susceptible to Rayleigh-Taylor instabilities which are not apparent here \citep{2009MNRAS.400..629P}. 

Stagnant dust structures have recently been reported in another WC+O CWB system: WR~112. The origin of the fragmented nested shells seen in WR~112 \citep{2017ApJ...835L..31L} was left as an open question, although the authors suggest that the structure may have arisen from previous periods of Roche lobe overflow. 

The most common physical picture to explain such puzzling systems invokes a ``switch in state'' from a past episode in which mass loss was governed by different physics, in particular slower winds. However, we show here that this idea is unlikely to work for Apep. Unlike WR~112, Apep does exhibit a precisely constrained expansion speed allowing the entire structure to be kinematically aged. Dust in the outermost coils of the nebula would have been ejected from the Central Engine about 125~years previously (see Supplementary Information section\,\ref{sec:diff}), while many of the innermost features are significantly younger ($\leq 50$\,yr). Under the hypothesis of a recent switch-in-state from slow to fast, the latter wind (now an order of magnitude faster) should rapidly overtake and collide with prior slow-wind structures. The timescale for this is very short —- the fast WR wind should sweep the entire volume to the edge of the visible plume in a little over a decade. One would expect such an event to disrupt the elegantly sculpted slow plume. Therefore, to invoke any kind of switch-in-state scenario like that suggested for WR~112, we must be witnessing a privileged moment where the switch has occurred very recently (within the last decade or so): a highly suspect coincidence.

With the fast and slow winds observationally confirmed (by spectroscopy and proper motion respectively), and with both manifest simultaneously in the Apep system, we instead propose that this duality of the wind must be intrinsic to the system. The most natural scenario in which all the evidence can be made to fit together is one in which the central engine WR star launches both a slow and fast wind. Anisotropic winds, for example configured as a fast polar and slow equatorial flow, have been established in other settings \citep{2006ApJ...638L..33G,2008A&A...485..245G} with rapid stellar rotation most often invoked as the underlying driver \citep{2004A&A...418..639A,2014A&A...562A.118S}. For the case of Apep, spectroscopic analysis of the SINFONI data, as well as the presence of the spiral plume, points to the central engine hosting an unresolved CWB, possibly WR-WR or WR-O (see Supplementary Information section\,\ref{sec:ir_spec_cent}). As the orbit of the binary companion takes it through the equatorial plane occupied by the slow heavy wind from the rapid rotator, a colliding-wind plume will result by way of the Pinwheel mechanism. Dust may continue to expand at the slow rate inflating the embedded spiral plume structure, unhampered by the fast polar wind with which it will never interact. This neatly ties together all the phenomenology, removing the conflict between spectroscopic and proper motion windspeeds, and implying a CWB orbital period comparable to the $\sim$125\,yr plume dynamical age or longer.

\begin{figure}
\begin{center}
    \includegraphics[scale=0.5]{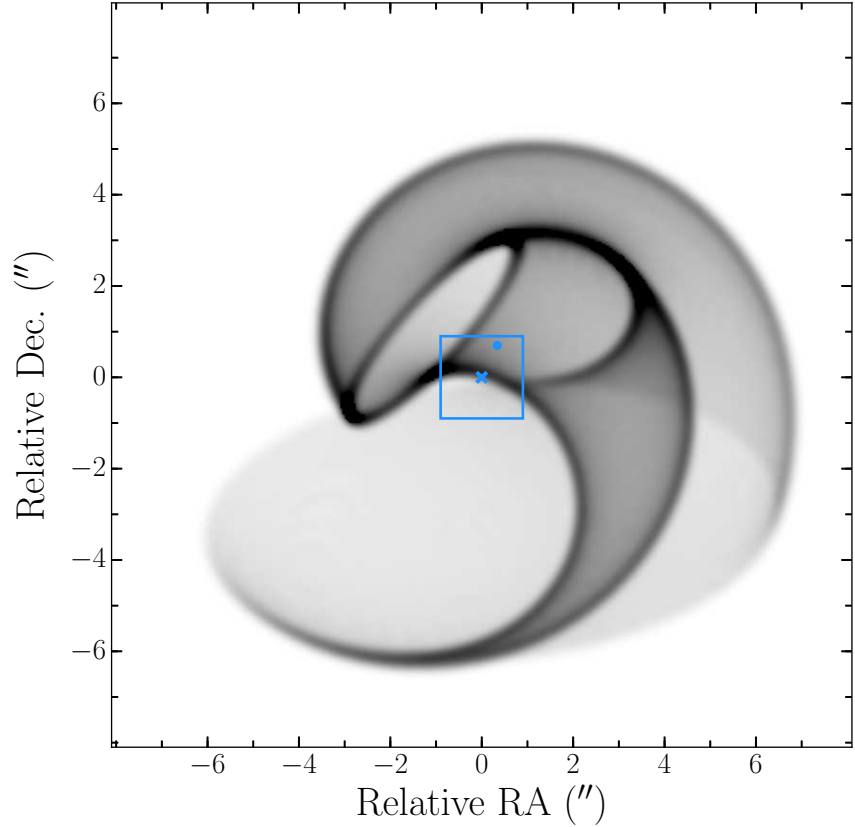}
    \caption{A geometric model of the dust plume of Apep. Note that the model only traces the surface of the colliding-wind dust plume. The mathematical origin of the Archimedian spiral (and the presumed location of the stars of the Central Engine) is indicated by the blue cross centered within the box. With this image registration, the location of the northern companion is also presented identically to Figure\,\ref{fig:visir_img}.}
    \label{fig:model_dust_plume}
\end{center}
\end{figure}

Although a detailed model optimized to fit all the structural elements and brightness profile of the plume lies beyond the scope of the present work, here we provide a simple plausible geometrical model to test our favored anisotropic mass loss channel. Our toy geometrical model for the Apep plume is shown in Figure\,\ref{fig:model_dust_plume} (with complete details provided about the model in Methods section\,\ref{sec:model_dust_plume}). In summary, Figure\,\ref{fig:model_dust_plume} results from a spiral with a full opening angle of $120^\circ$ (implying a nearly-equal momentum ratio of the colliding winds, which also favours a WR+WR composition of the Central Engine). The pole of the spiral is projected at $30^\circ$ to the line of sight which results in some overlapping structures. With the angular windspeed set to the measured value of 50\,mas\,yr$^{-1}$ (Supplementary Information section~\ref{sec:diff}) then the resultant image is produced with an orbital period of 130\,years -- essentially the same as the dynamical age of the plume.

Our simple model has done a remarkable job of producing an image with the same structural elements to those found in the observed infrared nebula (Figure~\ref{fig:visir_img}), and it has done so with few meaningful free parameters -- essentially just the cone opening angle, the inclination, and the CWB orbital period. However, there are two additional degrees of freedom in the model that need to be discussed. The model presented has a discrete ``turn-on'' and ``turn-off'' in dust production, so that no dust is formed over one-quarter of the binary period. The older (turn-on) event delimits the outer periphery of the plume; the model turn-off event implies a relatively abrupt halt in dust production about $\sim$60\,yr ago and it is this feature that produces the prominent inclined ellipse near the heart of the system.

Such an abrupt change in dust production state is part of the standard phenomenology of WR Pinwheels, and indeed almost exactly these kinds of shapes in the expanding plume are witnessed in the episodic dust producer WR~140 \cite{Williams2009}. However, we also note that the finding of abrupt changes in dust formation state would be expected under the assumption of our preferred dense slow equatorial wind model. In the (likely) event that the orbital plane of the companion star that is providing the colliding wind is not coincident with the plane of the equatorial wind, then strong dust production will only occur for that part of the orbit in which these intersect. Therefore, a plausible mechanism to switch dust production on and off in Apep naturally occurs as the colliding-wind companion passes through and emerges from the sector of dense slow wind from the equator of the rapid rotator.

In the event these lines of evidence pointing to an extreme anisotropic wind driven by rapid rotation in a WR host star are verified, the Apep system presents the first local astrophysical laboratory to study a potential long-duration gamma-ray burst progenitor. Angular momentum evolution through the transition to the Wolf-Rayet phase is poorly understood with varying mechanisms capable of yielding wildly divergent outcomes in rotation \citep{2014A&A...562A.118S}. Although, the most promising scenarios for spin-up entail close binary interaction or mergers discussed for the context of massive stars by de~Mink et al.\citep{deMink2013} Long-duration gamma-ray bursts are thought to originate from the core-collapse supernova of a WR star with high angular momentum, near its critical rotation rate \citep{1993ApJ...405..273W,2012A&A...547A..83G}. Apep may be the prototype for a new class of system of such stars, encouraging further detailed study to test GRB models that have largely been confronted only with post-explosion data at extragalactic distances.

\bibliographystyle{naturemag}

\section*{Acknowledgements}
J.~R.~C. thanks B. Gaensler and S. Farrell for conceiving the radio and X-ray survey that led to the discovery of Apep. The authors thank R. Lau (Caltech) and 2 other (anonymous) referees for comments and critique that lead to important improvements to this manuscript. We also thank N. Smith, K. Valenta, and O. de Marco for discussions in the early stages of this study, and A. Cheetham for help in constructing the NACO and VISIR observing schedules. P.~G.~T. and B.~J.~S.~P. are grateful for funding from the Breakthrough Prize Foundation. This work was performed in part under contract with the Jet Propulsion Laboratory (JPL) funded by NASA through the Sagan Fellowship Program executed by the NASA Exoplanet Science Institute. P.~M.~W. is grateful to the Institute for Astronomy for continued hospitality and access to the facilities of the Royal Observatory, Edinburgh. The authors acknowledge the Gadigal clan of the Eora nation, the traditional owners of the land on which the University of Sydney is built, and we pay our respects to their knowledge, and their elders past, present and future. The authors also thank Yinuo Han and Joshua Prenzler. This research has made use of the SIMBAD database, operated at CDS, Strasbourg, France, and NASA's Astrophysics Data System. This work has also made use of the \textsc{IPython} package \citep{PER-GRA:2007}; SciPy \citep{scipy};  \textsc{matplotlib}, a \textsc{Python} library for publication quality graphics \citep{Hunter:2007}; \textsc{Astropy}, a community-developed core \textsc{Python} package for astronomy \citep{2013A&A...558A..33A}; and \textsc{NumPy} \citep{van2011numpy}. The results presented in this Letter are based on observations collected at the European Organisation for Astronomical Research in the Southern Hemisphere under ESO programmes 097.C-0679(A), 097.C-0679(B), 299.C-5032(A), and 299.C-5032(B). The Australia Telescope Compact Array is part of the Australia Telescope National Facility which is funded by the Commonwealth of Australia for operation as a National Facility managed by CSIRO. The scientific results reported in this article are based in part on data acquired through the Australian Astronomical Observatory, on data obtained from the \emph{Chandra} Data Archive, and observations obtained with \emph{XMM-Newton}, an European Space Agency (ESA) science mission with instruments and contributions directly funded by ESA Member States and NASA.

\textbf{Authors contributions:} J.~R.~C. conducted the survey that identified Apep, wrote the initial draft of the manuscript, conducted and reduced the ATCA observations, and reduced the IRIS2, SINFONI, \emph{XMM-Newton}, and \emph{Chandra} observations. P.~G.~T. measured the proper motion of the dust spiral, led the discussion and interpretation of the object, led the VISIR and NACO observing proposals, and contributed significantly to the writing and editing of the manuscript. B.~J.~S.~P. contributed significantly to the understanding and discussion of the object, provided editing and text for the manuscript, and analysed the \emph{Gaia} and NACO data. P.~M.~W. interpreted the infrared spectrum, provided the text for the manuscript, produced the infrared/optical photometric SED, measured the equivalent widths of the emission lines, and contributed to the discussion about the object. P.~A.~C. helped interpret the infrared spectra and critiqued the manuscript. M.~E. and B.~N. reduced the VISIR and NACO data. L.~K-C. conducted the IRIS2 observation.

\textbf{Competing interests:} The authors have no competing interests with respect to this manuscript.

\textbf{Data and materials availability:} All data included in this manuscript are now available in the public domain. The VISIR, NACO, and SINFONI data are available through the ESO archive. The ATCA data are available through the Australia Telescope Online Archive (ATOA). The IRIS2 data are available through the AAT Data Archive. The X-ray data are available through the \emph{XMM-Newton} Science Archive (XSA) and \emph{Chandra} data archive. 

\newpage

\section{Methods}

\subsection{Observations and data reduction}

\subsubsection{NACO}

Apep was observed 2016 April 28 with the NACO camera \citep{2003SPIE.4841..944L,2003SPIE.4839..140R} on the VLT in three different filters, 2.24\,$\mu$m, 3.74\,$\mu$m, and 4.05\,$\mu$m. The 3.74\,$\mu$m and 4.05\,$\mu$m images were taken with the L27 camera with a pixel scale of 27.15 mas/pix, and the 2.24\,$\mu$m image was taken with the S13 camera, which has a scale of 13.22 mas/pix. The 4.05\,$\mu$m observations were taken using chopping and hardware windowing of the detector, with a chopping frequency of $\approx$\,0.1\,Hz. 

The NACO data were reduced via the standard ESO pipeline (v4.4.3). Dark subtraction, flat fielding, and sky subtraction were performed on the standard star and Apep. The sky frame for all filters were computed from frames that were taken 5-10 minutes either side of the Apep frame. The task $\tt{jitter}$ was used to produce the image data. Zero points for all filters were calculated using the standard star HD\,144648. Since two other 2MASS stars were detected in the 2.24\,$\mu$m observations, we corrected to the astrometric world coordinate system of 2MASS with a shift of $\approx$\,0.5$''$ vertically and $\approx$\,2$''$ horizontally. The resulting $K$-band image is shown in the inset of Figure\,\ref{fig:visir_img}.

We note that the 2.24\,$\mu$m image of the Central Engine shows that the source is slightly resolved. A two-dimensional isotropic Gaussian was fit to the visibilities as function of baseline, finding a best fit of $\sigma = 28\pm12$~mas, assuming 2\% visibility uncertainty. We do not obtain a satisfactory fit with a binary model. We expect that this measurement is dominated by systematics rather than the 12~mas statistical uncertainty, but it can be safely stated that the central engine is marginally resolved in the NACO images.

\subsubsection{VISIR}

Mid-Infrared imaging observations were made with the VISIR instrument \citep{2004Msngr.117...12L} on 2016 August 13 and 2017 August 1 in the 8.9\,$\mu$m filter, as well as on 2016 July 23 in the 11.7\,$\mu$m filter. All images had a plate scale of 45~mas/pix, and implemented a parallel chop-nod pattern with a chopping frequency of $\approx$\,4\,Hz. The 2016 VISIR data was reduced using the chopping and nodding method via the standard ESO developed pipeline (v4.3.1). Due to an incorrect setting in the observing block, the chopping amplitude for the 2017 data was somewhat too small, however the impact of this on the final data products was minor. Full imagery was recovered by judicious application of a window function, although this required some custom processing codes. The resulting 8.9\,$\mu$m of the 2017 August 1 epoch, after applying a high-pass filter, is shown in the left panel of Supplementary Information Figure\,\ref{fig:visir_diff}.

\subsubsection{SINFONI}

We observed Apep using the near-IR integral field spectrometer SINFONI instrument \citep{2003SPIE.4841.1548E} at the VLT in $J$- and $H$+$K$-bands on 2017 July 21. The nominal spectral resolution for SINFONI in $J$- and $H$+$K$-band is 2000 and 1500, respectively. Apep was observed for a total $\approx$\,25 minutes in each band, with nodding performed every 30\,s and 3\,s for $J$- and $H$+$K$-band, respectively. A plate scale of 0.1$''$\,pixel$^{-1}$ were used for both filters, providing a $3'' \times 3''$ field-of-view. The observations were adaptive-optics assisted using a natural guide star. We observed the standard star HIP\,082670 to correct for spectral telluric features. 

The standard ESO SINFONI pipeline (v3.0.0) via \emph{Gasgano} was used to perform the data reduction. Dark subtraction, flat fielding, detector linearity, geometrical distortion, and wavelength calibration were applied to Apep, the standard star, and sky frames. Sky subtraction was then performed via the standard pipeline \citep{2007MNRAS.375.1099D}. 

The standard star's spectrum was extracted from a data cube using an aperture three times larger than the PSF to ensure all the flux was captured. The spectrum was then normalised by a black body curve of the appropriate temperature. Intrinsic spectral features of the standard star were removed through modelling of the lines with Lorentzian profiles before correcting for atmospheric transmission curves in the science data cubes.

Apep was resolved by SINFONI into two sources in both $J$- and $H$+$K$-band. We summed the $J$-band data over the Central Engine and northern companion to derive the $J$-band magnitudes of 10.2\,$\pm$\,0.2 and 9.6\,$\pm$\,0.2, respectively. 

\subsubsection{IRIS2}

On 2017 March 16, we observed Apep with the long-slit spectrograph IRIS2 \citep{2004SPIE.5492..998T} on the 3.9\,m AAT at the Siding Spring Observatory, Australia. The IRIS2 instrument is based on a HAWAII1 HgCdTe detector, and can achieve a spectral resolution of $ \approx 2400$. Our observation was conducted using the $J_{s}$-filter, which is sensitive between 1.041 and 1.256\,$\mu$m. Exposure times were 18\,s on Apep before nodding Apep to another position along the slit. A total of eights scans were taken, resulting in a total of $2.5$\,h on source. HIP\,73881 was the standard star observed to correct for atmospheric transmission features, of which ten scans of 18\,s length were taken. The IRIS2 slit has a width of 1$''$on sky, implying that the long-slit spectrum observed is the combination of both sources identified in the NACO image presented in the inset of Figure~\ref{fig:visir_img}. 

The data reduction of $J_{s}$-band used the standard routines in the Image Reduction and Analysis Facility (IRAF, v2.16) software. Firstly, the spectra were flat-fielded and dark corrected. Spectra were extracted, as well as sky subtracted, for each scan via the task {\tt apall}. Wavelength calibration was done using a Xe arc lamp spectra. Each scan of the standard star was combined using the task {\tt scombine}, and the Paschen-$\gamma$ absorption feature 1.094\,$\mu$m was removed via fitting a Lorentzian profile. After combining all the science spectra, the telluric features were removed using the task {\tt telluric}. To remove the presence of the northern companion from the 1.083\,$\mu$m\,He\,I line, the aperture used to extract the spectrum was adjusted to isolate the broad component, which corresponds to emission from the WR star in the Central Engine \citep{1994MNRAS.269.1082E}. The line with the contamination of the northern companion removed is presented in Figure~\ref{fig:he1_line}.

\subsubsection{ATCA}

We observed Apep on 2017 May 11 with the ATCA in the 6A array configuration, which facilitates the highest resolution imaging of any ATCA configuration. The observations were conducted at the central frequencies of 2.1 and 17/19\,GHz, with total integration times of 180 and 200 minutes, respectively. The Compact Array Broadband Backend (CABB) \citep{Wilson2011} provides instantaneous 2\,GHz bandwidth for both linear polarizations at all frequencies. Each observation used a 10 second correlator integration time and 1\,MHz channels. Antenna-1 (CA01) was not available during the observations due to maintenance activities.

The calibrators PKS~B0823-500 and PKS~B1921-293 were used for flux density and bandpass calibration of the 2.1 and 17/19\,GHz observations, respectively. PKS~B0823-500 is known to be slightly variable, introducing a 10\% systematic uncertainty in the flux density scale in the 2.1\,GHz-centred observation. The flat-spectrum source PMN~J1534-5351 was the phase calibrator for all frequencies. PMN~J1534-5351 was observed after 30 minutes on source for frequencies below 10\,GHz. For the 17/19\,GHz observations, the phase calibrator was targeted after 20 minutes on source, and pointing calibration was performed every 50 minutes using PKS~B1921-293.

The data was reduced using the \textsc{miriad} software package \citep{Sault1995}. Initially, we flagged areas of known radio frequency interference (RFI) and lower sensitivity in CABB. Excision of RFI from the flux density and bandpass calibrators was performed using the automatic flagging option in the task $\tt{pgflag}$, and then manually with $\tt{blflag}$. For the observation centred on 2.1\,GHz, $\approx$\,30\% of the data was flagged. For all the other observations, $\approx$\,10\% of the data was flagged. The instantaneous 2\,GHz bandwidth was split into four 512\,MHz wide subbands, in which bandpass, gain, and leakage solutions were estimated using the flux density calibrator and ten second time intervals. The calculated calibration solutions were transferred to the phase calibrator PMN~J1534-5351. RFI excision was similarly performed before calculating the phase solutions and ensuring the flux density of PMN~J1534-5351 was bootstrapped to the flux density scale defined by the gain calibrator. All calibration solutions were then transferred to Apep.

We applied multi-frequency synthesis over the 512\,MHz bandwidth of each subband to image Apep with a robust parameter of $-1.0$. The images were \textsc{clean}ed to the first negative \textsc{clean} component before going through a phase self-calibration step using the shallowly \textsc{clean}ed images as a model. After re-imaging the self-calibrated data, the images were \textsc{clean}ed more deeply to approximately three times the theoretical rms level. The resulting rms noise in for the 1.4 and 19.7\,GHz images varied from $\approx$\,1 to $\approx$\,8\,$\times 10^{-2}$\,mJy\,beam$^{-1}$, respectively. The contours of the 19.7\,GHz image is shown in the inset of Figure\,\ref{fig:visir_img}.

\subsubsection{Chandra and XMM-Newton}

Apep was observed between 0.2 and 10.0\,keV on six different occasions by the \xmm~and \chan~X-ray observatories. The \chan~X-ray observatory detections and two of the \xmm~observatory detections of Apep were recorded serendipitously, since Apep is located within 10$'$ of the well-studied supernova remnant G330.2+1.0. The observations are described in greater detail in Supplementary Information section\,\ref{sec:xrays}. The counts of the X-ray observations make Apep the fourth brightest CWB thus far detected \citep{2018MNRAS.474.3228P}.

The \xmm~data were reduced following the standard procedure \citep{Callingham2012}. The \xmm~Science Analysis System (version 16.0.0) software was used to ingest the observation data files and to produce the source spectra. The European Photon Imaging Camera (EPIC) data was filtered for background flares.

We selected appropriate spectral extraction regions depending on the off-axis angle of Apep. Background spectra were extracted from the same CCD as the source spectra with annular areas two times larger than the source extraction region. To minimise the impact of pile-up, only single events were included in the production of the spectra. Photon redistribution matrices and ancillary response files were constructed for the spectra, which were than grouped into 20 counts per energy bin before being read into XSPEC (version 12.9.1) \citep{Arnaud1996} for model fitting. 

The \chan~X-ray observatory observations were reduced using the \chan~Interactive Analysis of Observations (CIAO, version 4.9) software and the calibration database CALDB (version 4.7.5.1). In all \chan~observations, Apep was detected with the Advanced CCD Imaging Spectrometer (ACIS)-I3 at a large off-axis angle $>8'$. Time-dependent gain correction, the latest map of the ACIS gain, and a correction for charge transfer inefficiency were all applied to the data. The data was filtered for flaring and bad event grades.

For spectroscopy, response and effective area files were produced using the CIAO task {\tt specextract}. The resulting source and background spectra were also grouped in 20 counts per energy bin and converted into a file format that could be read into XSPEC. 

\subsection{Model of the dust plume} \label{sec:model_dust_plume}

Our simple plausible geometric model allows us to draw preliminary connections between physical conditions at the Central Engine driving the flow and the morphology of the plume from infrared imagery. Even without a detailed fit, such a toy model is able to provide a useful bridge: confronting the data with expectations from an idealized spiral outflow.

Our model assumes dust is entirely distributed on the surface of a cone, with the opening angle determined from the momentum ratio of the colliding winds (see e.g. Tuthill et al.\cite{tuthill2008}). Treated as an optically thin surface which becomes more dilute (faint) in proportion to its radial displacement from the central host star along the outflow, this cone is wrapped into an Archimedean spiral by the orbit of the binary in the Central Engine -- following the commonly understood underlying model for a Pinwheel. Apart from the cone opening angle, most physical quantities governing the apparent form of a classical Pinwheel (such as WR~104) are dictated by the orbital parameters of the central binary. Most of these free parameters largely just alter the orientation of the object on the sky and will not be discussed further. Important parameters to highlight are the inclination of the binary orbit to the line of sight (which can dramatically change the appearance), and the product of the windspeed and the orbital period which governs the winding angle of the spiral pattern.

As discussed in the main body, our toy geometrical model for the Apep plume, presented in Figure~\ref{fig:model_dust_plume}, results from a spiral with a full opening angle of $120^\circ$ (implying a nearly-equal momentum ratio of the colliding winds, which also favours a WC7+WN4-5 composition of the Central Engine). The pole of the spiral is projected at $30^\circ$ to the line of sight which results in some overlapping structures. With the angular windspeed set to the measured value of 50\,mas\,yr$^{-1}$ (from Section~\ref{sec:diff}) then the resultant image is produced with an orbital period of 130\,years.

The model does require discrete ``turn-on'' and ``turn-off'' phases in dust production, so that no dust is formed over one-quarter of the binary period.  The older (turn-on) event delimits the outer edge of the plume; without it the model would continue with successive expanding coils of dust. Although this turn-on may well have happened in Apep, it is also possible that the outer periphery of the dust is simply too cool, tenuous or possibly eroded by UV to be seen, so that this parameter may be considered a proxy for these enabling our simple model to provide a cosmetic match. This is not the case for the inner turn-off event. The model turn-off event implies a relatively abrupt halt in dust production about $\sim$60\,yr ago and it is this feature that produces the prominent inclined ellipse near the heart of the system. This apparent ``cavity'' just North-East of the core is one of the more noteworthy and puzzling features seen in the original infrared imagery.

Such an abrupt switch in dust production state is part of the standard phenomenology of WR Pinwheels, and indeed almost exactly these kinds of shapes in the expanding plume are witnessed in the episodic dust producer WR~140 \cite{Williams2009}. For the WR~140 CWB system, periodic sharp rises in infrared flux indicative of short episodes of active dust formation are entirely governed by the proximity of the companion star in an 8\,yr eccentric orbit. A similar mechanism might also be invoked for Apep, although the toy plume model here, which requires dust formation over most of the orbital cycle, would seem not to support this idea. As highlighted in the main text, in the event the companion star's orbital plane is roated relative to the plane of the equatorial wind, then strong dust production will only occur for that part of the orbit in which these intersect. Therefore, we can achieve a natural switch in dust production as the companion passes through the dense slow wind produce by a rapidly rotating primary.

\newpage

\section{Supplementary Information} 
\setcounter{figure}{0} 

\subsection{Differential image and measuring the proper motion of the dust spiral} \label{sec:diff}

As noted in the main body text, proper motions witnessed over the 353 day ($\sim$1\,year) interval between the two VISIR epochs were about one order of magnitude smaller than anticipated. Simple differential imaging revealed changes consistent with a pure radial expansion of the plume between epochs, however the modest size of the motions required further image processing to extract quantitative estimates of the displacement. Firstly the reduced and cleaned VISIR images from both epochs were highpass filtered with a 20-pixel (0.9$''$) Gaussian kernel, accentuating edges and boundaries of the plume (see the left panel of Supplementary Information Figure~\ref{fig:visir_diff}). The expansion over the elapsed time interval may be observed in a differential highpass image as shown in the right panel of Supplementary Information Figure~\ref{fig:visir_diff}. The result is consistent with a uniform isotropic expansion of the astrophysical structures over time, but not with noise processes such as a change in seeing between the intervals. 

\begin{figure*}[h]
\renewcommand\figurename{Supplementary Information Figure}
\begin{center}$
\begin{array}{cc}
\includegraphics[scale=0.4]{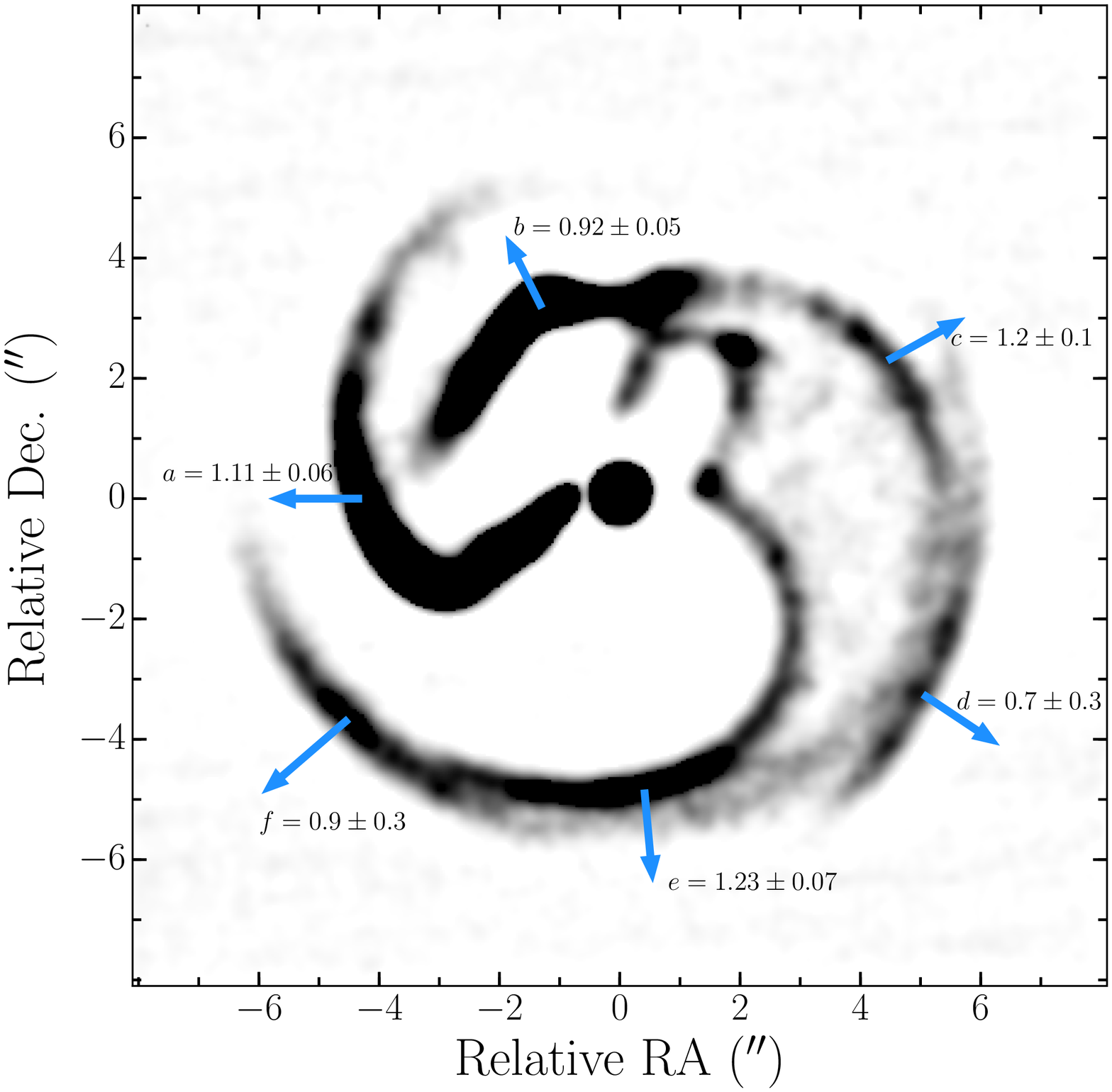}&
\includegraphics[scale=0.4]{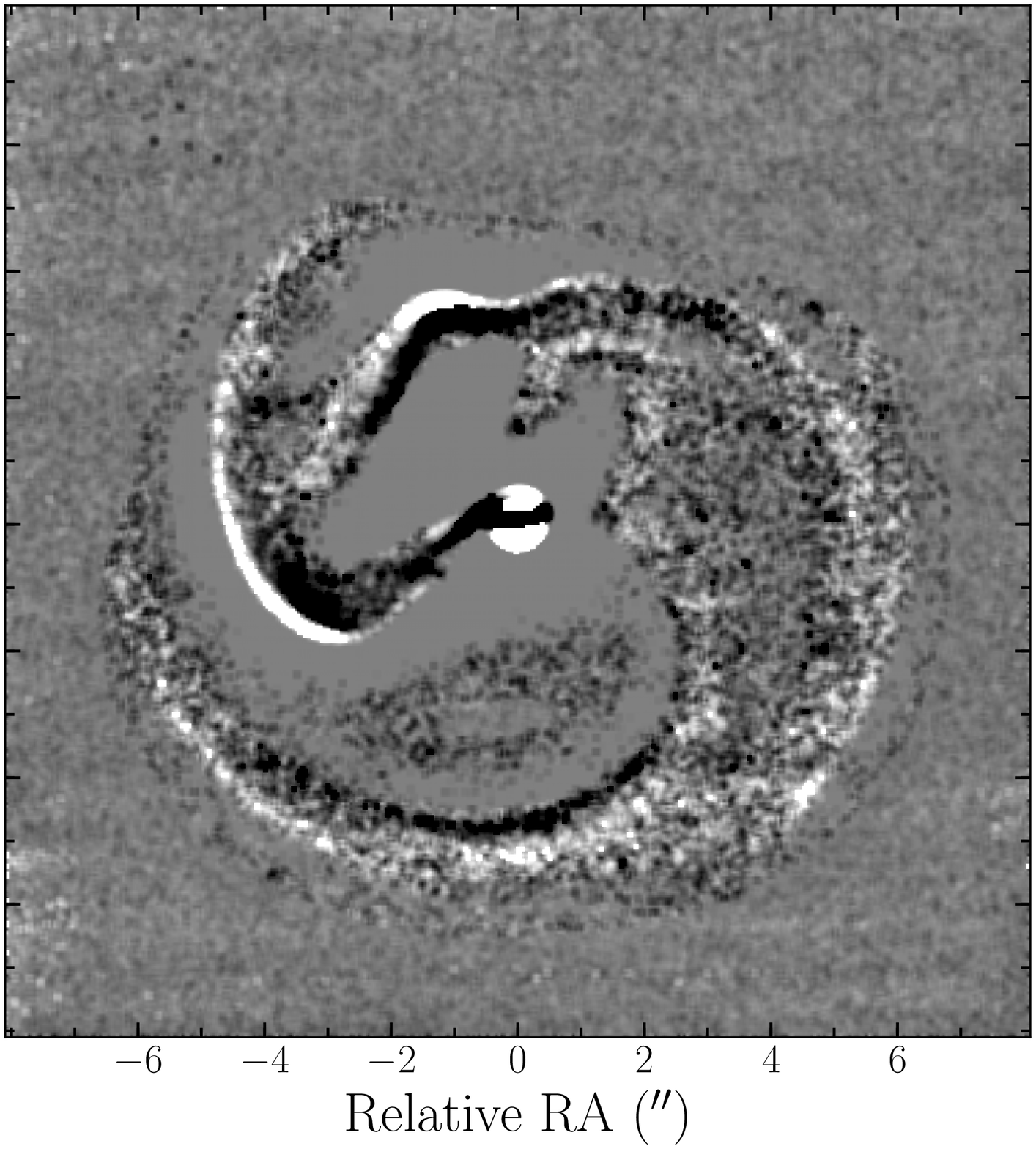}
\end{array}$
 \caption{A highpass filtered image of the 1 August 2017 8.9\,$\mu$m Apep VISIR data emphasizing the location of edges and boundaries of the plume structure (left panel), and an illustration of the proper motion recorded over the $\sim$year interval between the two VISIR epochs provided by subtracting the 2016 from the 2017 highpass filtered images (right panel). Left panel: The blue arrows indicate features labelled a-f with annotations recording measured displacement in pixels between the 2016 and 2017 epochs. Note that the pixel scale is 45 mas. The uncertainties on the pixel rates are for 90\% confidence. Right Panel: The real displacement in the edges and boundaries demarking the dust over the interval is witnessed as the newer epoch (white) is found exterior to the older epoch (black).}
\label{fig:visir_diff}
\end{center}
\end{figure*}

Although the differential image has established the presence of inflation of the structure, accurate measurement of the radial windspeed required further analysis. This proceeded by fitting a piecewise linear curve to the ridge crests in the highpass filtered images at each epoch separately. In Supplementary Information Figure~\ref{fig:visir_diff}, data following the six most prominent arcs (a--f corresponding to edges in the original image presented in Figure~\ref{fig:visir_img}) along the majority of their length were extracted. The radial displacements of these data points were then registered against each other over the two epochs, and the displacement of each arc averaged over its length. It was found that these highpass filtered edges, which trace the major structural elements of the plume, were all found to exhibit positive radial displacements between 2016 and 2017. Some spread in recorded windspeed values is expected as most likely not all the edges are moving exactly in the plane of the sky, resulting in some apparent slowing due to projection. Countering this effect, limb brightening through an optically thin structure will preferentially yield edges which lie close to the plane of the sky, and so projection is unlikely to be a critical factor. 

To get a quantitative estimate of the expansion speed, we took the mean of the pixel displacement measured from the structures labeled $a$, $b$, $c$, and $e$ in the left panel of Supplementary Information Figure\,\ref{fig:visir_diff}. These four structures have the highest signal-to-noise ratio for the displacement measurement, are roughly evenly distributed azimuthally around the dust plume, and are presumably the least projected features. The average of the measured displacement of these four structures provides an average angular measurement of 50\,$\pm$\,6\,mas\,yr$^{-1}$. At a distance of 2.4\,kpc, this corresponds to an expansion velocity of 570\,$\pm$\,70\,km\,s$^{1}$. 

Given such a well defined (and seemingly constant) outflow speed everywhere around the plume, we are able to estimate the dynamical age of any feature in dust assuming an origin at the central engine. Most of the outermost spiral features are at about 6.3\,arcseconds away, implying an ejection from the core 125\,$\pm$\,15 years ago.

\subsection{Distance and possible open cluster association}
\label{sec:dist}

Due to the presence of diffuse interstellar bands (DIBs) in the spectra of Apep, we can place an upper limit on the distance to the system. The detected DIB at 1.527\,$\mu$m has an equivalent width of $1.02 \pm 0.04$\,\AA. Based on the relationship between the equivalent width of the 1.527\,$\mu$m DIB line and foreground extinction $A_{\mathrm{v}}$, this implies the distance to Apep is $\lesssim 4.5$\,kpc \citep{2015ApJ...798...35Z}.

If we use the visual extinction $A_{V} = 11.4$ estimated from the broadband photometric SED (see Supplementary Information section\,\ref{sec:opt_ir_sed}) and the known visual magnitude $V = 17.5$ for Apep ($V = 17.8$ for the OB supergiant that is the northern companion and $V = 19.0$ for the Central Engine) we need a distance of $d = 2.4^{+0.2}_{-0.5}$ to get realistic absolute magnitudes for the components \cite{2015MNRAS.447.2322R}. We note that there are large number of uncertainties in all the above measured quantities, particularly $A_{V}$ and the absolute magnitudes of the components.  

The X-ray and radio luminosities also suggest the system should not be significantly further away than 2.4\,kpc. As alluded to in the main text, if we were to put the system at a distance of $\approx$\,9\,kpc to make the proper motion velocity of the dust spiral consistent with the spectroscopically measured wind velocity, the 0.2-10\,keV X-ray and 1.4\,GHz radio luminosity of Apep would be $L_{0.2-10\,\mathrm{keV}} \approx 2 \times 10^{35}$\,erg\,s$^{-1}$ and $ L_{1.4\,\mathrm{GHz}} \approx 3 \times 10^{31}$ \,erg\,s$^{-1}$, respectively. Such luminosities would make Apep the brightest peristent X-ray CWB in the Galaxy \citep{2018MNRAS.474.3228P}, and an order of magnitude brighter in radio emission than $\eta$~Carinae, the most luminous known radio CWB (for direct comparison to Table\,A.5 of De Becker et al.\citep{2013A&A...558A..28D}, the radio luminosity of Apep, integrated over the bandpass of 0.33 to 30\,GHz, is $L_{\mathrm{rad}} \approx 2 \times 10^{32}$\,erg\,s$^{-1}$ at this distance). 

If we place the system at 4.5\,kpc, corresponding to the upper limit derived for the distance to Apep, we still have more than a factor of three discrepancy between the velocity of the dust spiral and the spectroscopically measured wind velocity, and luminosities of $L_{0.2-10\,\mathrm{keV}} \approx 5 \times 10^{34}$\,erg\,s$^{-1}$ and $ L_{1.4\,\mathrm{GHz}} \approx 6 \times 10^{30}$ \,erg\,s$^{-1}$ ($L_{\mathrm{rad}} \approx 5 \times 10^{31}$\,erg\,s$^{-1}$). At this luminosity Apep would now be the third most luminous CWB in X-rays \citep{2018MNRAS.474.3228P} and with a similar radio luminosity to $\eta$~Carinae when not in outburst \citep{2013A&A...558A..28D}.

Despite these uncertainties, all lines of evidence suggest that Apep is located $\lesssim 4.5$\,kpc, and likely around $d \approx 2.4$\,kpc. We also note that the position of Apep is co-incident with the proposed young open cluster Majaess 170, which has a dust extinction of $E(J - H) = 0.38 \pm 0.03$ \cite{2013Ap&SS.344..175M,2017MNRAS.464.1119C}. However, such an association is likely co-incidental since the dust extinction of Apep  $E(J - H) \sim 1$, derived from the 1.527\,$\mu$m DIB line \cite{2016MNRAS.463.2653D}, is significantly different. Furthermore, this cluster is not apparent in either distance or proper motion as measured by \emph{Gaia}\citep{2016A&A...595A...2G,BailerJones18}, and we suggest that Apep likely lies behind the cluster. 

\subsubsection{Gaia}

In its Data Release~2, the ESA~\emph{Gaia} Mission \citep{2016A&A...595A...2G}  delivered parallaxes for more than a billion stars, including the Central Engine of Apep but not the slightly brighter northern companion. The \emph{Gaia} source 5981635832593607040 is identified with the Central Engine of Apep, and has a parallax of $1.78 \pm 0.29$~mas, significantly higher than would be expected for an object at $\sim 2$~kpc. Correcting these for systematics and population statistics\citep{BailerJones18} yields a distance of $575_{-89}^{+129}$~pc for the Central Engine, much closer than the photometrically derived distance quoted above. This very short distance would drop the absolute $V$~band magnitude of the OB~supergiant northern companion from $\sim -6.1$ at 2.4~kpc to $\sim -3.3$, and the~WN/WC+O or WN+WC Central Engine from $\sim -4.8$ to $\sim -2.1$, which in both cases would mean dropping from a reasonable luminosity for a star of its type to far below what is expected \cite{2015MNRAS.447.2322R}.

We believe that the Gaia distance measurement is likely to be a systematically underestimated relative to the true distance of Apep due to the central engine binary motion affecting the \emph{Gaia} astrometry and, probably more importantly, the hierarchical triple arrangement of the system. 

The central engine has a \emph{Gaia}~DR2 `astrometric excess noise' parameter  of 1.24~mas and the northern companion of 5.40~mas, both at high significance, especially the latter. This parameter represents the amount of error that has to be added in quadrature to the formal parallax uncertainty in the astrometric solution, determined by nonlinear optimization similar to the common approach of adding error to set reduced $\chi^2$ to unity, to represent additional noise sources such as binary motion and blending \citep{lindegren12}. The values for the northern companion and Central Engine of Apep are both much higher than for ordinary field stars, and indeed significantly larger than the formal 0.29~mas parallax uncertainty, and excess noise parameters greater than 1~mas can be indicative of binary motion perturbing the parallax fit \citep{lindegren18,evans18}. We also note that despite the northern companion (\emph{Gaia} source 5981635832568990208
) being 0.9 magnitudes brighter than the Central Engine in the \emph{Gaia} $G$-band, no parallax measurement is reported.

\emph{Gaia} pixels are elongated, with a scale of 59~mas on the short axis and 177~mas on the long axis. Faint \emph{Gaia} sources are observed with pixels binned along their long axes in strips of~12 for an effective long axis scale of 2.12''. Therefore, in scans perpendicular to our hierarchical triple the northern companion and Central Engine are well-resolved, but in scans parallel to the binary axis they are confused, biasing any parallax measurement.

We therefore suggest that the \emph{Gaia} distance to Apep must unfortunately be discounted in this analysis, as it is likely subject to significant systematic error. We hope that the longer timebase, astrometric binary orbits, and deblending planned to be performed in \emph{Gaia}~DR4 (expected~2022) will provide an accurate parallax to the Central Engine and northern companion of Apep.

\subsection{Spectral classifications of the Central Engine}
\label{sec:ir_spec_cent}

The SINFONI spectra of the Central Engine (Figure~\ref{fig:sinfoni_spec}, Table~\ref{tab:line_id_rex1a}) show C\,III-IV and He\,II emission lines characteristic of a WC type WR star. The diagnostic C\,IV/C\,III 1.191/1.199\,$\mu$m line ratio is close to 3.0, which matches that for WC7 stars \citep{2018MNRAS.473.2853R} and is significantly greater than the 1.4 in the WC8 standard WR~135 \citep{1991MNRAS.252..300E}, indicating the likely presence of a WC7 type star in the Central Engine of Apep. The presence of a late-type WC star is also supported by the observed far-infrared colour excess in WISE, MSX \citep{Egan1996}, and AKARI \citep{AKARI_IRC,AKARI_FIS} flux measurements of Apep (see Supplementary Information section\,\ref{sec:opt_ir_sed}).

\begin{table*}
\renewcommand\tablename{Supplementary Information Table}
  \small
  \caption{\label{tab:line_id_rex1a} Wavelengths ($\lambda$), identifications, and equivalent widths (EWs) of the lines identified in the SINFONI and IRIS2 spectra of the Central Engine. EWs of blends are given for the first contributor only, with others marked `(bl)'. The uncertainties on the EWs represent 3-$\sigma$, and do not account for blending.}
  \begin{center}
    \begin{tabular}{lll}
      \hline
      $\lambda$ ($\mu$m) & Line ID   & EW ({\AA}) \\
       \hline
       \hline 
      1.083 & He I 2p-2s     &   87 $\pm$ 7  \\
      1.163 & He II (7-5)    &  156 $\pm$ 6 \\
      1.191 & C IV (8-7)     &   59 $\pm$ 6 \\
      1.199 & C III 4p-4s    &   19 $\pm$ 3 \\
      1.256 & C III (9-7)    &   17  $\pm$ 3 \\
      1.282 & He I (5-3)   &   56 $\pm$ 6 \\
      1.282 & He II (10-6)   & (bl)\\
      1.476 & He II (9-6)  &   36 $\pm$ 3\\
      1.476 & C IV (18-12)   & (bl)\\
      1.552 & O V (10-9)   &    4 $\pm$ 2 \\
      1.572 & He II (13-7)  &   36 $\pm$ 5 \\
      1.575 & C III (13-9)   & (bl)\\
      1.693 & He II (12-7)   &   29 $\pm$ 4 \\
      1.701 & He I 4d-3p   & (bl)\\
      1.736 & C IV (9-8)     &   31 $\pm$ 3 \\
      2.078 & C IV 3d-3p     &   43  $\pm$ 5 \\
      2.108 & C III 5p-5s    &   26  $\pm$ 3 \\
      2.113 & He I 4s-3p     & (bl)\\
      2.117 & C III (8-7)    & (bl)\\
      2.189 & He II (10-7)   &   25  $\pm$ 3 \\
    \hline\end{tabular}
\end{center}
\end{table*}

However, the spectrum of Apep shows stronger He\,II and weaker C\,IV line emission than is stereotypical for a WC7 star. The emission line weakness is dilution by additional continuum, partly from dust emission, which increases with wavelength: e.g. the ratio of the He\,II 2.189/1.163-$\mu$m lines is 0.16, compared with an average $\sim 0.38$ in other WC7 stars. The weakness in the $J$-band, where dust emission is negligible (top panel; Figure~\ref{fig:sinfoni_spec}), points to the additional continuum from a companion star. The abnormal strength of the He~II lines for a WC7 star suggests an early WN sub-type companion. The absence of N~V and relative weakness of He~I, and with comparison to WN spectra \citep{1996A&A...305..541C}, implies the presence of a subtype WN4 or WN5 star. Double WR binaries are, however, rare, with very few known\citep{2004ApJ...611L..33B}.

An alternative spectral subtype classification to the WC7+WN4-5 model, that equally well describes the spectra shown in Figure~\ref{fig:sinfoni_spec}, is that of a WR star in the brief transitory phase between WN and WC (WN/WC) with an unseen OB-type companion. The WN/WC classification accurately describes the line ratios of the C\,III, C\,IV, and He\,I lines, and the abnormal strength of He\,II lines \citep{2018MNRAS.473.2853R}. However, WN/WC stars have never been observed as dust producers, or as part of a Pinwheel nebula, despite several known binary systems \citep{1989ApJ...344..870M,2009MNRAS.399.1977M}. One way to discern between a double WR binary or a WN/WC+O composition is through the detection of the 0.971\,$\mu$m C\,III line. If the Central Engine is a WN/WC+O, the 0.971\,$\mu$m C\,III line would be relatively weak compared to the Helium lines, whereas the line would be strong for a double WR binary.

We favour the WC7+WN4-5 model for the Central Engine of Apep, on the basis that no transitional WN/WC star has been observed to be a dust producer or part of a Pinwheel nebula, possibly on account of their low carbon abundances \citep{2012A&A...540A.144S}. In comparison, the prolific dust maker WR48a \citep{2012MNRAS.420.2526W} appears to be a WC8+WN8h \citep{2014MNRAS.445.1663Z}. In either event, spectroscopy of the Central Engine shows it to be a CWB, with at least one component a WR star. Independent of whether the binary is a double WR binary or a WN/WC+O, in both situations we would expect a wind an order of magnitude larger than that observed in the proper motion of the dust spiral seen in the VISIR data of Apep.

\subsection{Spectral properties of the distant northern companion}

The SINFONI spectra of the northern companion are presented in Supplementary Information Figure~\ref{fig:sinfoni_spec_rex1b}. The near-infrared spectrum is dominanted by Paschen-$\beta$, Brackett-$\gamma$, and 1.083\,$\mu$m He\,I (as shown by the IRIS2 data) line emission. In particular, the $K$-band spectrum resembles the O8Iaf supergiant HD151804 \citep{1999ApJ...511..374B}. However, the spectrum lacks the required 2.189\,$\mu$m He\,II line of an O81af supergiant. Lacking any He\,II lines, the spectrum resembles the B1\,Ia+ supergiant HD 169454 \citep{2007A&A...465..993G}. However, that would imply a reasonably large high luminosity for a B1\,Ia+ supergiant, with an absolute magnitude of -9.2. Despite this, we favour the northern companion being an B1\,Ia+ supergiant but further observations, particularly optical spectra, are necessary to confirm this spectral type. We note that the 1.083\,$\mu$m He\,I and Paschen-$\beta$ lines display a P~Cygni profile, both providing a measurement of a terminal windspeed of $\approx$\,900\,km\,s$^{-1}$. 

\begin{figure*}[h]
\renewcommand\figurename{Supplementary Information Figure} 
\begin{center}$
\begin{array}{cc}
\includegraphics[scale=0.4]{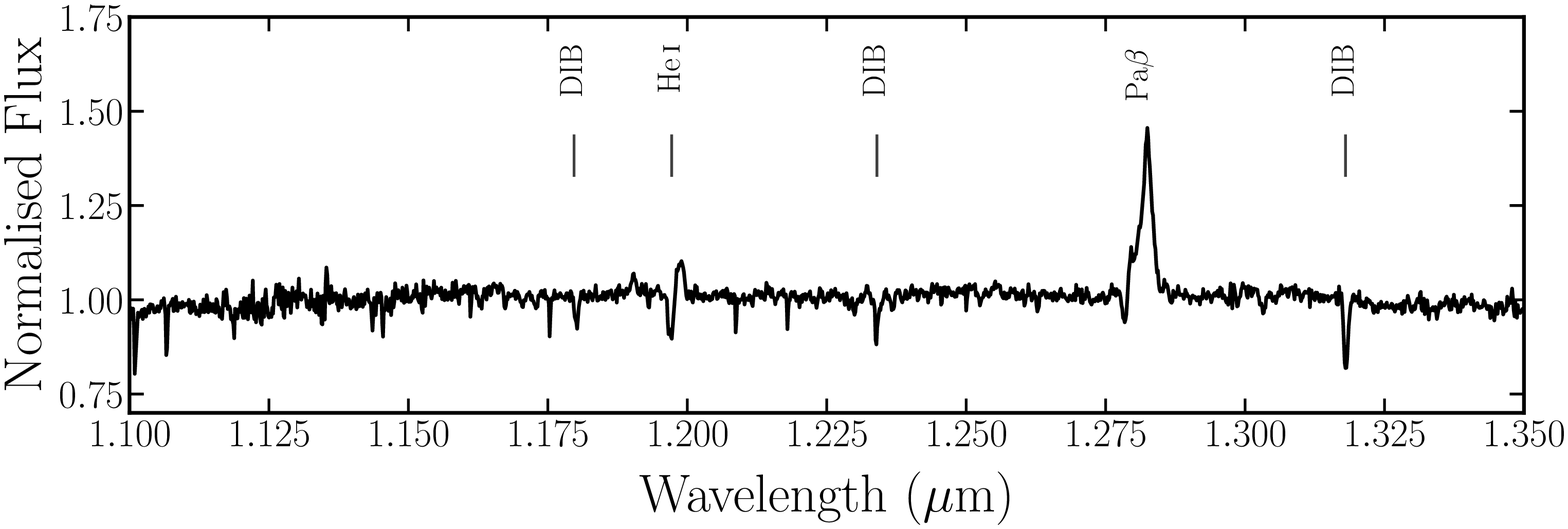}\\
\includegraphics[scale=0.4]{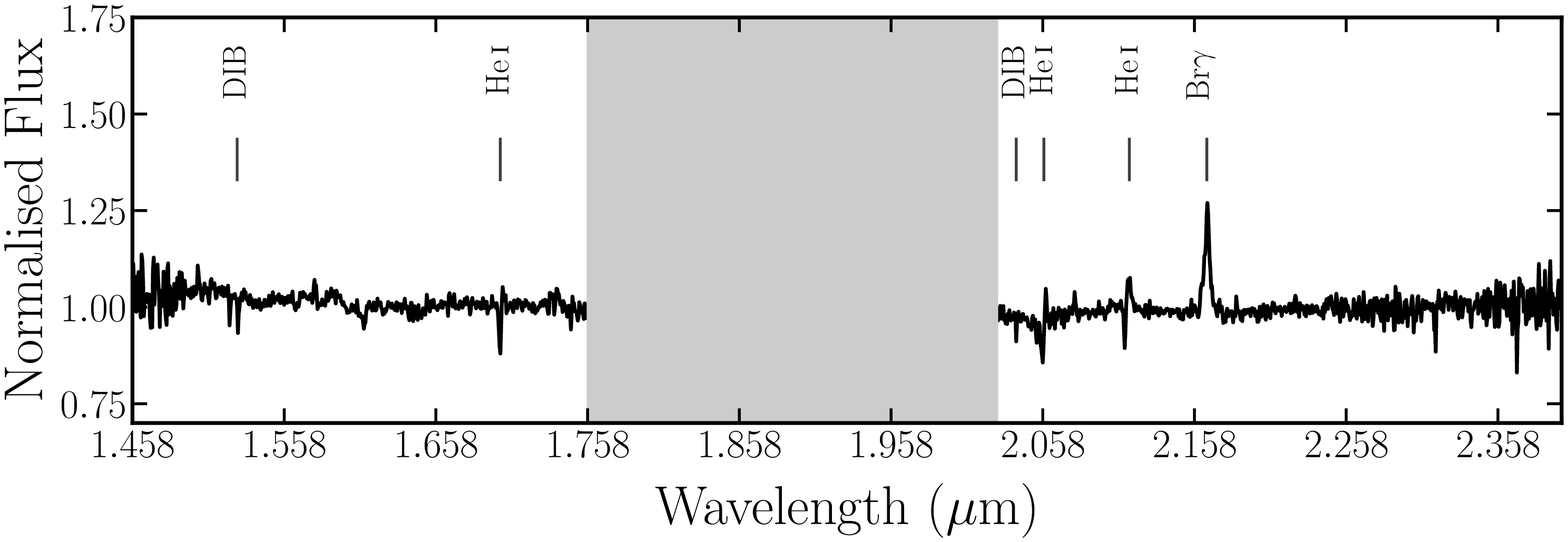} \\
\end{array}$
 \caption{SINFONI $J$-band (top) and $H$+$K$-band (bottom) spectra for the northern companion. Prominent emission lines are labelled and the area where telluric correction was not possible in the $H$+$K$-band is indicated in gray. Known and suggested diffuse interstellar bands are labelled by `DIB'.}
\label{fig:sinfoni_spec_rex1b}
\end{center}
\end{figure*}

\subsection{Optical and infrared photometry of Apep}
\label{sec:opt_ir_sed}

The SED derived from the infrared and optical photometry of Apep is provided in Supplementary Information Figure\,\ref{fig:sed_ir}. Our NACO observations, magnitudes, and angular separation of the Central Engine and northern companion are summarised in Supplementary Information Table\,\ref{tab:naco}. While our SINFONI and NACO data yield $J$-, $K$-band, 3.74- and 4.05-$\mu$m photometry of the Central Engine and northern companion separately, all other data are for the system as a whole. Sources of the short wavelength photometry are the DECam Plane Survey \citep{DECamPS} for $g$ and $r$, the VPHAS+ survey \citep{VPHAS} for $r$ and $i$, the DENIS survey \citep{DENIS} for $i$ (at a slightly longer wavelength, and treated separately for reddening and conversion to monochomatic fluxes), and the VVV survey \citep{2010NewA...15..433M} for $Z$. Infrared data for the system were taken from the 2MASS \citep{Skrutskie2006} survey for $J$, $H$ and $K_s$, the AllWISE release from the WISE mission for $W1$, $W2$, $W3$, and $W4$, \emph{Spitzer} GLIMPSE for the [5.8]-band, the AKARI IRC All-Sky \citep{AKARI_IRC} and FIS \citep{AKARI_FIS} surveys for $S9W$, $L18W$ and $N60$, the MSX SPIRIT III \citep{MSX_GP} for Bands $A$, $C$, $D$ and $E$ (8.28--21.3\,$\mu$m), and the \emph{Herschel} Hi-GAL survey \citep{HiGAL} for the 70\,$\mu$m flux. All four {\em WISE} bands were affected by saturation so we used profile-fitted magnitudes, which gave good matches to the central source in $W1$ and $W2$ but failed to capture all the extended emission in $W3$. The $W4$ magnitude included extended emission ($\chi ^2 = 124$) and matches the MSX flux at a close wavelength. Unfortunately, the {\em Spitzer} MIPSGAL 24-$\mu$m image was severely saturated. Although GLIMPSE did not give an 8-$\mu$m magnitude, we measured [8.0] = 0.33 from an {\em Spitzer} IRAC frame observed in AOR 20272896 (P.I. Benjamin) retrieved from the archive. The flux from this is also plotted in Supplementary Information Figure\,\ref{fig:sed_ir}.

\begin{table}
\renewcommand\tablename{Supplementary Information Table}
  \small
  \caption{\label{tab:naco} Summary of the NACO observations of Apep. Separation refers to the angular separation between the Central Engine and northern companion, identified in the inset of Figure~\ref{fig:visir_img}. The uncertainties reported are for 90\% confidence.}
  \begin{center}
    \begin{tabular}{llll}
    \hline
Wavelength (band) & Separation & Central Engine  & Northern companion   \\
   ($\mu$m)         &    ($''$)    &  (magnitude)     &    (magnitude)   \\
    \hline    
    \hline    
2.24 ($K$)    & 0.739\,$\pm$\,0.002      & 6.9\,$\pm$\,0.2          & 8.1\,$\pm$\,0.2    \\
3.74 ($L$)    & 0.745\,$\pm$\,0.008    & 4.7\,$\pm$\,0.1           & 7.3\,$\pm$\,0.1     \\
4.05 ($M$)    & 0.75\,$\pm$\,0.01        & 4.4\,$\pm$\,0.3           & 7.0\,$\pm$\,0.2    \\
    \hline\end{tabular}
\end{center}
\end{table}

\begin{figure}
\renewcommand\figurename{Supplementary Information Figure}
\begin{center}
    \includegraphics[scale=0.6]{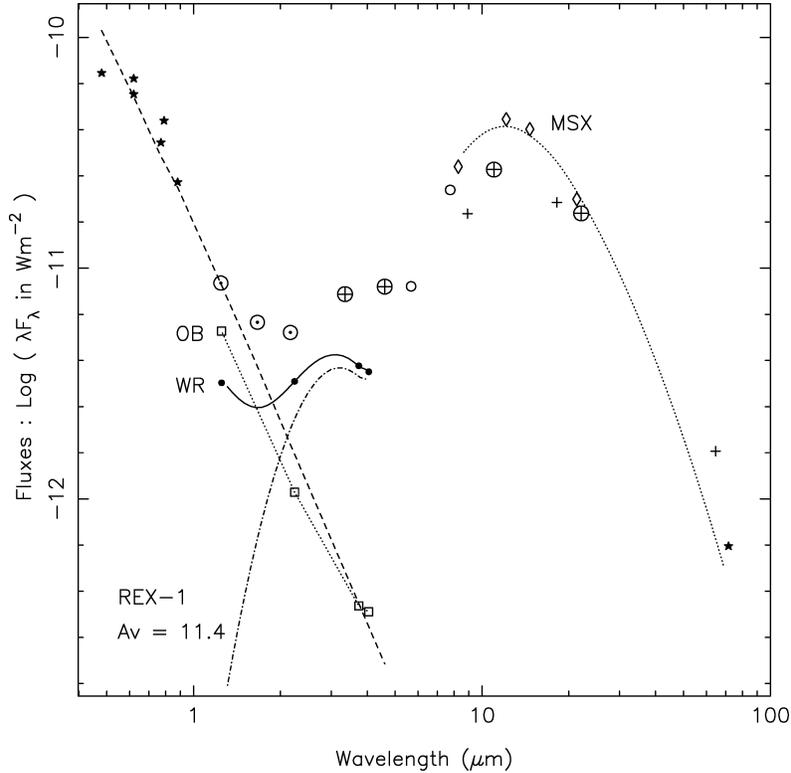}
    \caption{De-reddened optical to far-infrared SED of Apep. Fluxes were computed from $g$, $r$, $i$ and $Z$ from the DECam Plane (marked $\star$), VPHAS+, DENIS and VVV surveys, $J$-, $H$- and $K_s$-band from 2MASS (marked $\odot$), $W1$, $W2$, $W3$, and $W4$ from WISE (marked $\oplus$), [5.8]-band from GLIMPSE and our IRAC [8.0] (marked $\circ$), $S9W$, $L18W$, $N60$ and $WIDE-S$ from AKARI (marked +), MSX $A$, $C$, $D$ and $E$ (marked $\Diamond$), and the 70-micron Hi-GAL flux (marked $\star$). The dashed line fitted to the $g-Z$ points represents the SED of a 25\,kK supergiant, adopted for the northern companion and assumed to dominate the SED at these shorter wavelengths. Fluxes for the individual Central Engine and northern companion measured from our SINFONI and NACO observations are plotted $\bullet$ (`WR') and $\Box$ (`OB'), respectively. The latter, joined by a dotted line, are consistent with a stellar continuum, while the former (joined by a solid line) are modelled with the sum of a WR wind continuum fitted to $J$-band plus heated dust; the SED of that dust is also plotted dash-dot. We note that the VISIR 8.9 and 11.7 micron fluxes are consistent with MSX $A$ and the dotted fit.}
    \label{fig:sed_ir}
\end{center}
\end{figure}

Although the photometric data are heterogeneous in observation date and their coverage of the extended dust emission, we can use them for a robust measure of the interstellar reddening, the distance to the system, and to probe the dust formation history of Apep. 

The reddening was determined to be $A_V = 11.4$ by fitting the dust-free ($griZ$) photometry to the output flux of a WM-Basic \citep{WMBasic} model for a 25\,kK solar composition supergiant\footnote{\url{http://zuserver2.star.ucl.ac.uk/~ljs/starburst/BM_models/index.html}}. This model was selected for the northern companion, which is assumed to dominate the SED at shorter wavelengths since it is 0.56~magnitude brighter in $J$-band, and brighter still at shorter wavelengths. For the interstellar reddening, we used the `Wd1+RCs' law \citep{Damineli}, duly adjusted for the wavelengths of photometric bands used in the present study. 

The Apep photometric data were observed over a long period, from 1996 (DENIS) to 2016 (DECam), and their reasonable consistency suggests that the stellar flux from Apep was not significantly variable over this interval. The $J$-band flux is seen to lie on the optically fitted SED (Supplementary Information Figure\,\ref{fig:sed_ir}), indicating that there is no significant contribution by dust emission at this wavelength.

It is evident that the form of the IR SED is quite unlike sterotypical dust making WR stars, which are generally Planckian  peaking near 3~$\mu$m, and which can be modelled using the smoothly varying carbon dust emissivity and Planck functions having temperatures $T_g$ falling off with distance $d$ from the stars as \citep{WHT} $T_g \propto d^{-0.4}$. Instead, the Apep SED rises to peak near 12~$\mu$m, which is well defined by the MSX values. The MSX A and C--E bands have effective beam sizes of $\approx$\,21--26$''$, implying practically all of the flux from the extended dust cloud observed with VISIR should be captured by the MSX photometry. A simple model for the dust emission, an isothermal dust cloud fitted to the MSX data, yields $T_g$ = 230~K and $(2.2\pm0.3) \times 10^{-5} M_{\odot}$ of amorphous carbon grains using optical constraints\citep{ZubkoC}. The error on the derived dust mass is a formal error from the least-squares fit and is dwarfed by the systematics, notably the adopted distance to Apep as the derived dust mass is proportional to the square of the distance. Assuming a dynamical timescale similar to the orbit of Apep of 125 years, this implies a dust production rate of $\approx 2 \times 10^{-7} M_{\odot}~\mathrm{yr^{-1}}$, which is similar to the dust production rates of other known pinwheels with WC members \cite{Williams2009}. The SED is analogous to the typical WR dust clouds but cooler, consistent with the dispersion of the dust made in a earlier epoch. The flux from this model is also plotted in Supplementary Information Figure\,\ref{fig:sed_ir}. The WISE $W4$, {\em Spitzer} IRAC 8-$\mu$m and \emph{Herschel} Hi-GAL 70-$\mu$m fluxes are seen to consistent with it, suggesting that the mass of cool dust in this `reservoir', and the stellar flux heating it, have not changed significantly since the MSX observations in 1996. The {\em MSX} $A$ band covers a great deal more of the astronomical `silicate' feature than the IRAC [8.0], so that the closeness of the fluxes from the two observations suggests that silicate emission, e.g. from an earlier evolutionary phase, is not significant in the spectrum of Apep.

For more recent dust formation, we can also model the SED of the WR component using the fluxes from our SINFONI and NACO observations. We assume that there is no dust emission contribution in the $J$-band and model the WR wind flux with a power law of index -2.97 \citep{Morris} and subtract this from the observations. The difference spectrum is fitted by an isothermal dust cloud having $T_g$ = 830~K and mass $(2.2\pm0.1) \times 10^{-8} M_{\odot}$ of carbon. This dust mass is in line with those determined for many WC7--9 stars \citep{WHT} and almost four orders of magnitude less than that in the extended cloud. As for the extended cloud, the derived dust mass depends on the square of the adopted distance to Apep, but the ratio of dust masses does not. These results suggest that it is possible that the dust formation rate of Apep has not been constant on a long (centuries) time-scale, but was significantly greater a $\sim$century ago, consistent with the CWB orbital derived from the plume's dynamical age.

\subsection{X-ray properties of Apep}
\label{sec:xrays}

The X-ray observatory, epoch, observatory observation identification numbers, off-axis angle of Apep during the X-rays observations, and the available X-ray instruments during the observation, are provided in Supplementary Information Table\,\ref{tab:Xrayobs}. The folded spectra for the X-ray observations of Apep are plotted in Supplementary Information Figure~\ref{fig:xray_seds}. The spectra display the combination of heavy absorption with the features of a collisionally-ionised thermal plasma. For example, a strong emission line near 6.7\,keV is present in all observations. Non-thermal X-ray emission in CWBs is expected to make an insignificant contribution to the overall stellar X-ray spectrum \cite{2010MNRAS.403.1657P}, so only thermal X-ray models were considered. We find the best fitting model is provided by a single photoelectrically absorbed optically-thin plasma model ({\tt apec}), consistent with other known X-ray bright CWBs that all require winds with speeds $> 1000$\,km\,s$^{-1}$ to reproduce the required X-ray luminosities \citep{Anderson2011}. During the fitting, the abundance parameter was allowed to vary. We do not get an improved fit for multi-temperature thermal plasma models, such as {\tt apec + apec}. We also attempted to fit the spectra with a non-equilibrium collisionally ionised shock ({\tt vpshock}) but got unrealistic shock ionisation ages. The best fitting parameters derived from the {\tt apec} model for each observation are presented in Supplementary Information Table\,\ref{tab:x_ray_fits}. 

\begin{table*}
\renewcommand\tablename{Supplementary Information Table}
  \small
  \caption{\label{tab:Xrayobs} Summary of the 0.2 and 10.0\,keV observations of Apep. ObsID corresponds to the unique identification number assigned to each observation by the respective X-ray observatory. $t_{\rm{eff}}$ is the total effective exposure time of the observation after filtering for background flaring.}
  \begin{center}
    \begin{tabular}{llllll}
    \hline
Observatory & Epoch & ObsID & Off-axis angle & Instrument & $t_{\rm{eff}}$   \\
            &       &       & ($'$)&           & (s)  \\
    \hline    
    \hline    
\xmm  & 2004 Aug 10 & 0201500101 & 0.6 & pn   & 8140 \\
    &           &            &     & MOS1 & 8480 \\
    &             &            &     & MOS2 & 8500 \\
\chan & 2006 May 21 & 6687       & 8.6 & ACIS-I3 & 49970 \\
\xmm  & 2008 Mar 20 & 0500300101 & 9.3 & MOS1 & 51900 \\
    &             &            &     & MOS2 & 52870 \\
\xmm  & 2015 Mar 08 & 0742050101 & 8.8 & pn   & 105600  \\
      &             &            &     & MOS2 & 136940  \\
\chan  & 2017 May 02 & 19163 & 8.2 & ACIS-I3   & 74140 \\
\chan  & 2017 May 05 & 20068  & 8.2 & ACIS-I3   & 77320  \\
    \hline\end{tabular}
\end{center}
\end{table*}

\begin{figure*}
\renewcommand\figurename{Supplementary Information Figure}
\begin{center}$
\begin{array}{cc}
\includegraphics[scale=0.3]{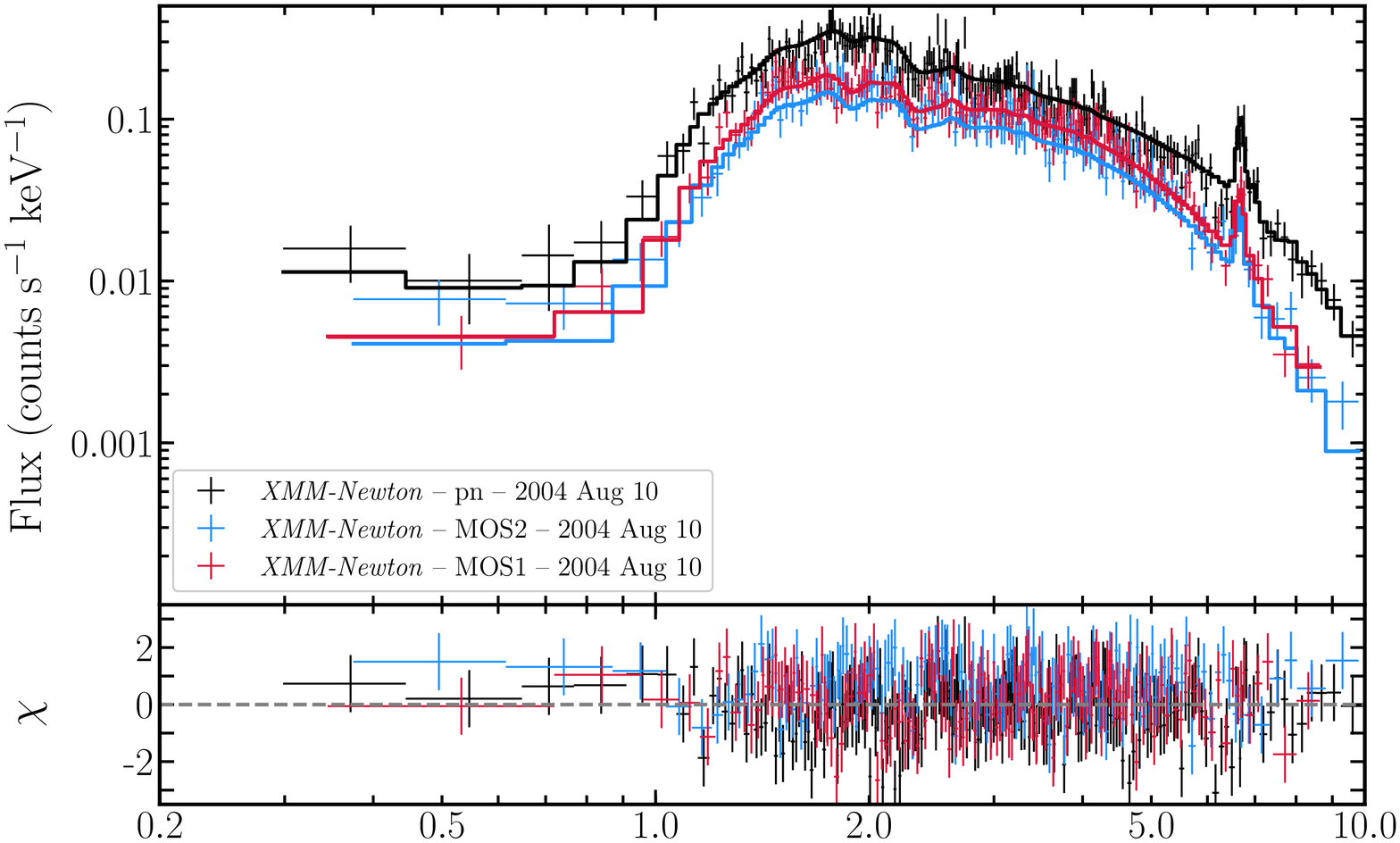} &
\includegraphics[scale=0.3]{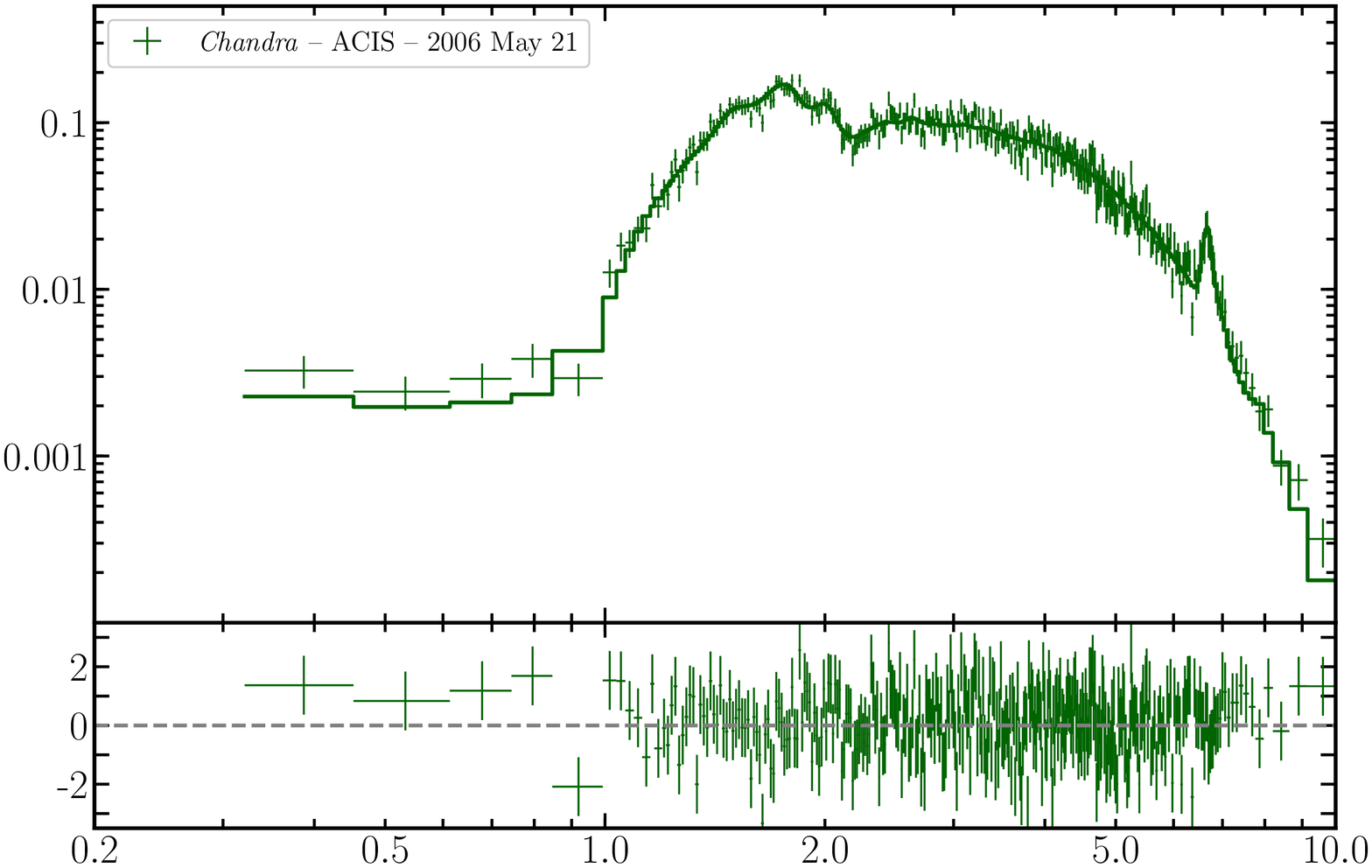}\\
\includegraphics[scale=0.3]{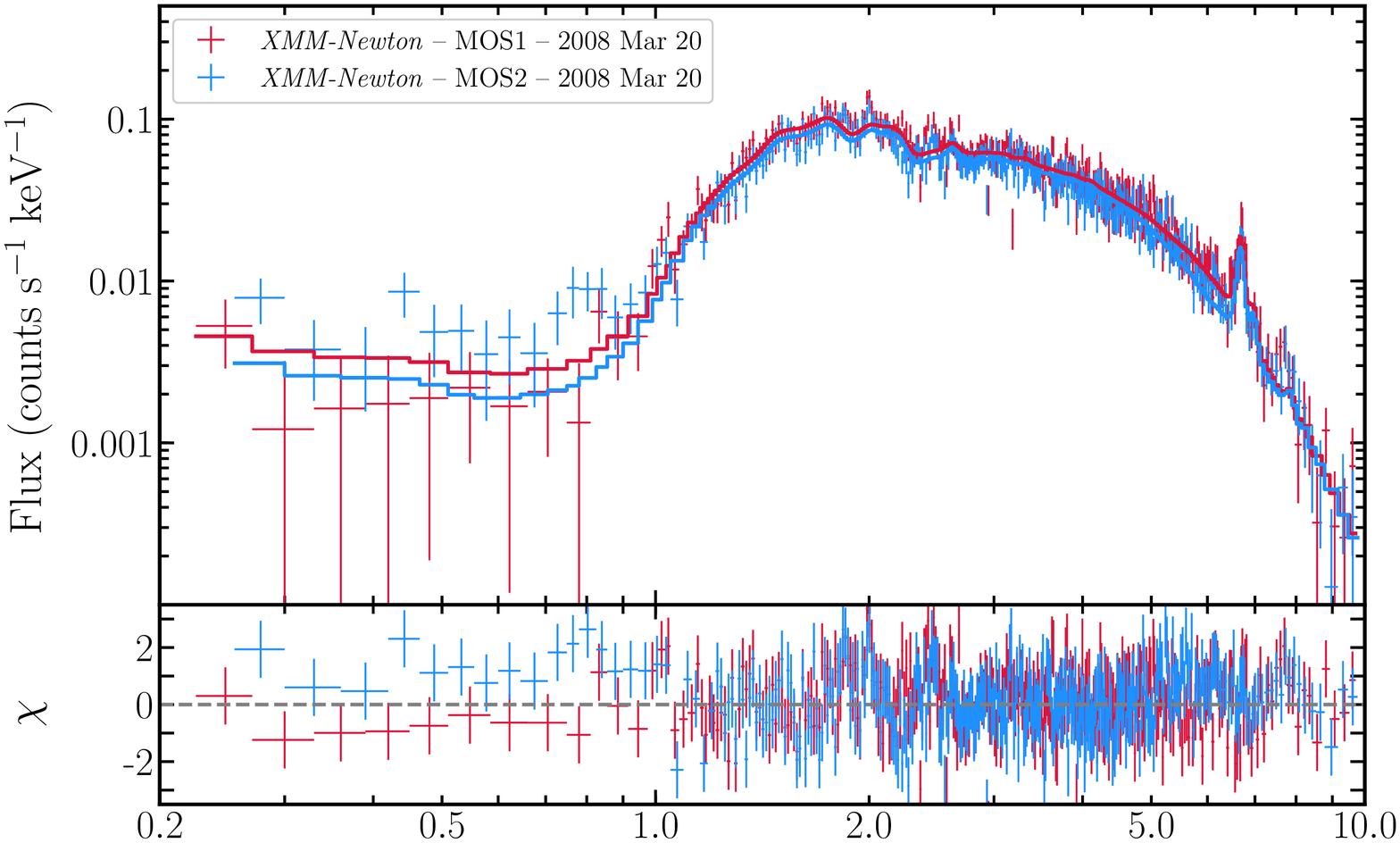} &
\includegraphics[scale=0.3]{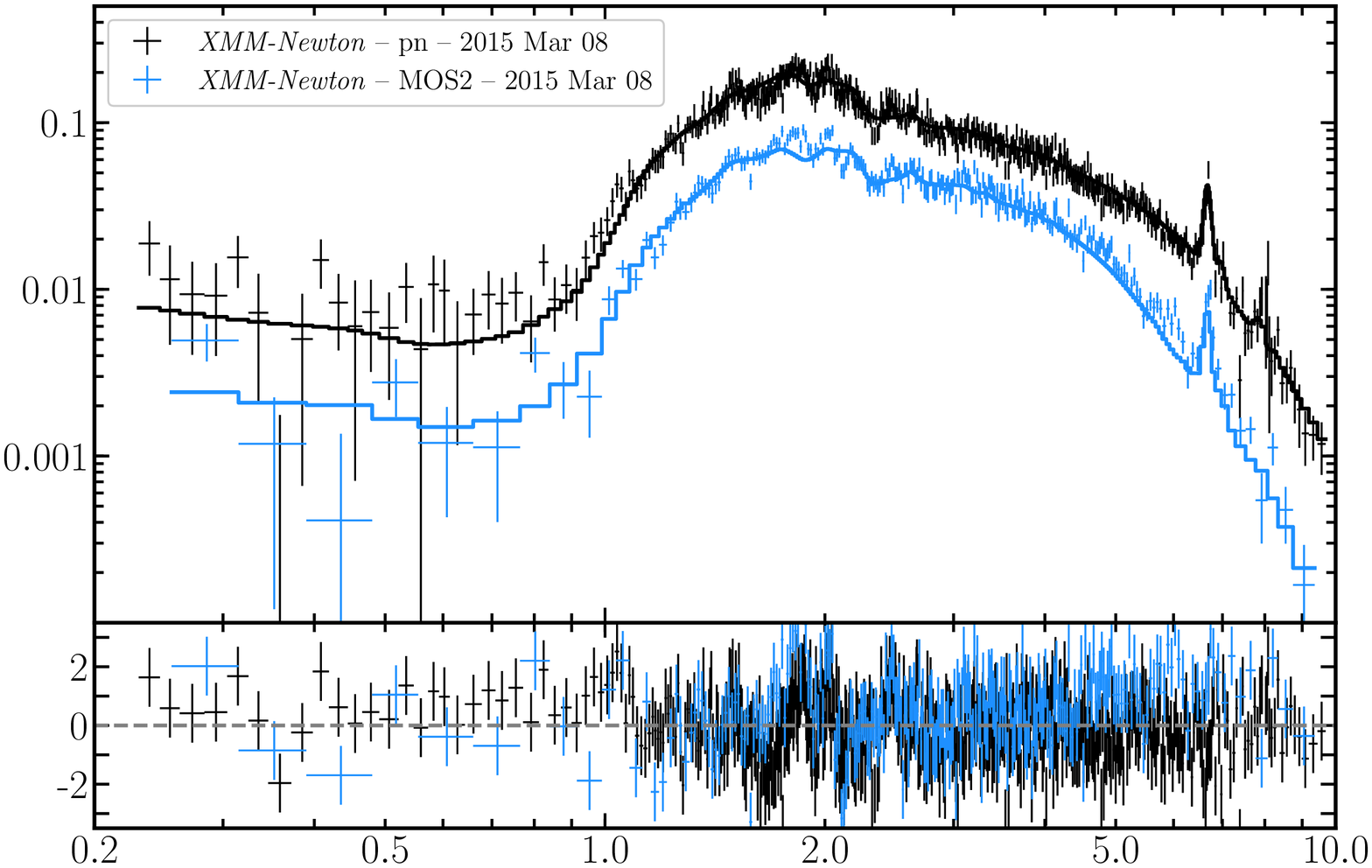} \\
\includegraphics[scale=0.3]{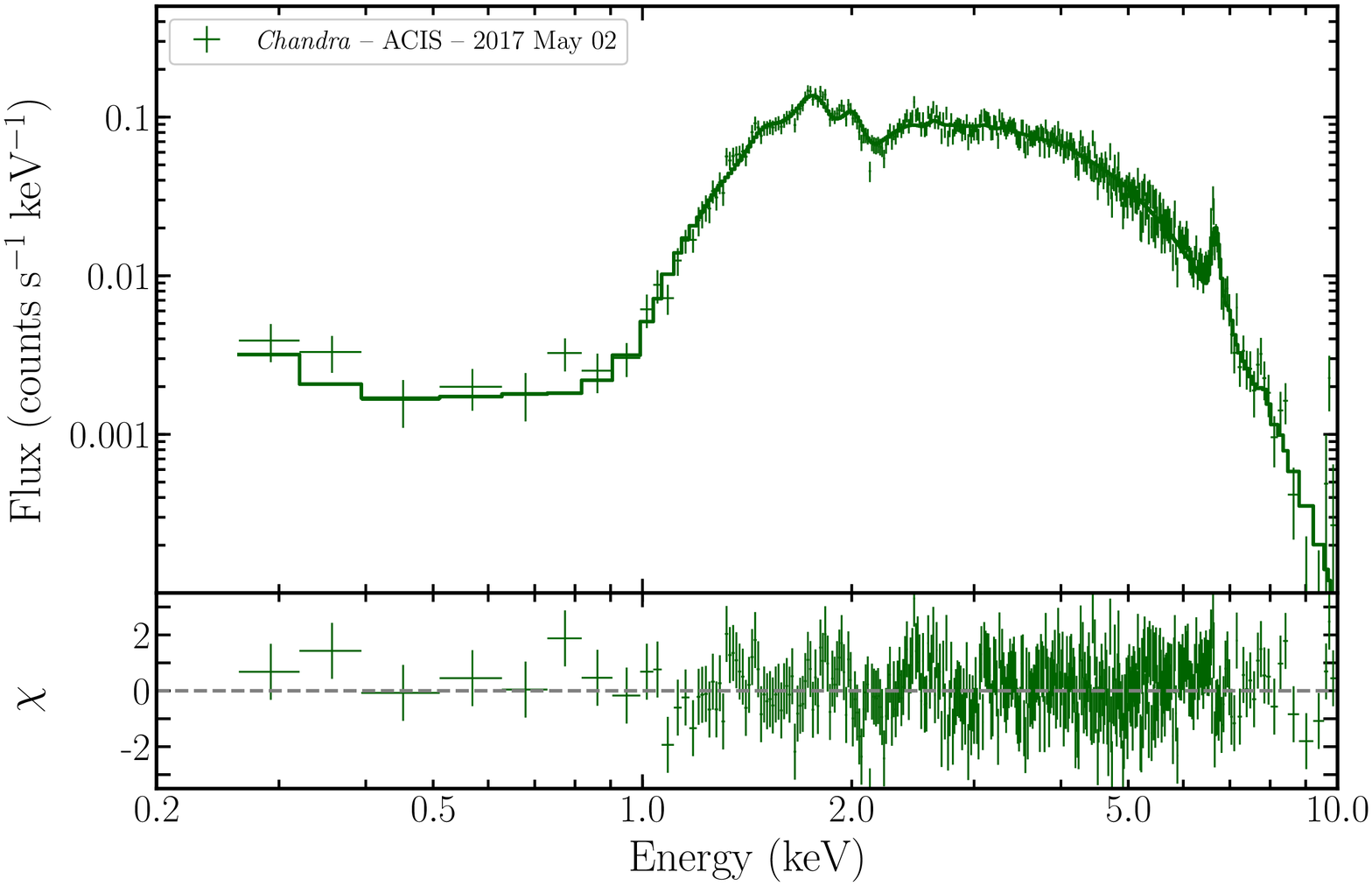} &
\includegraphics[scale=0.3]{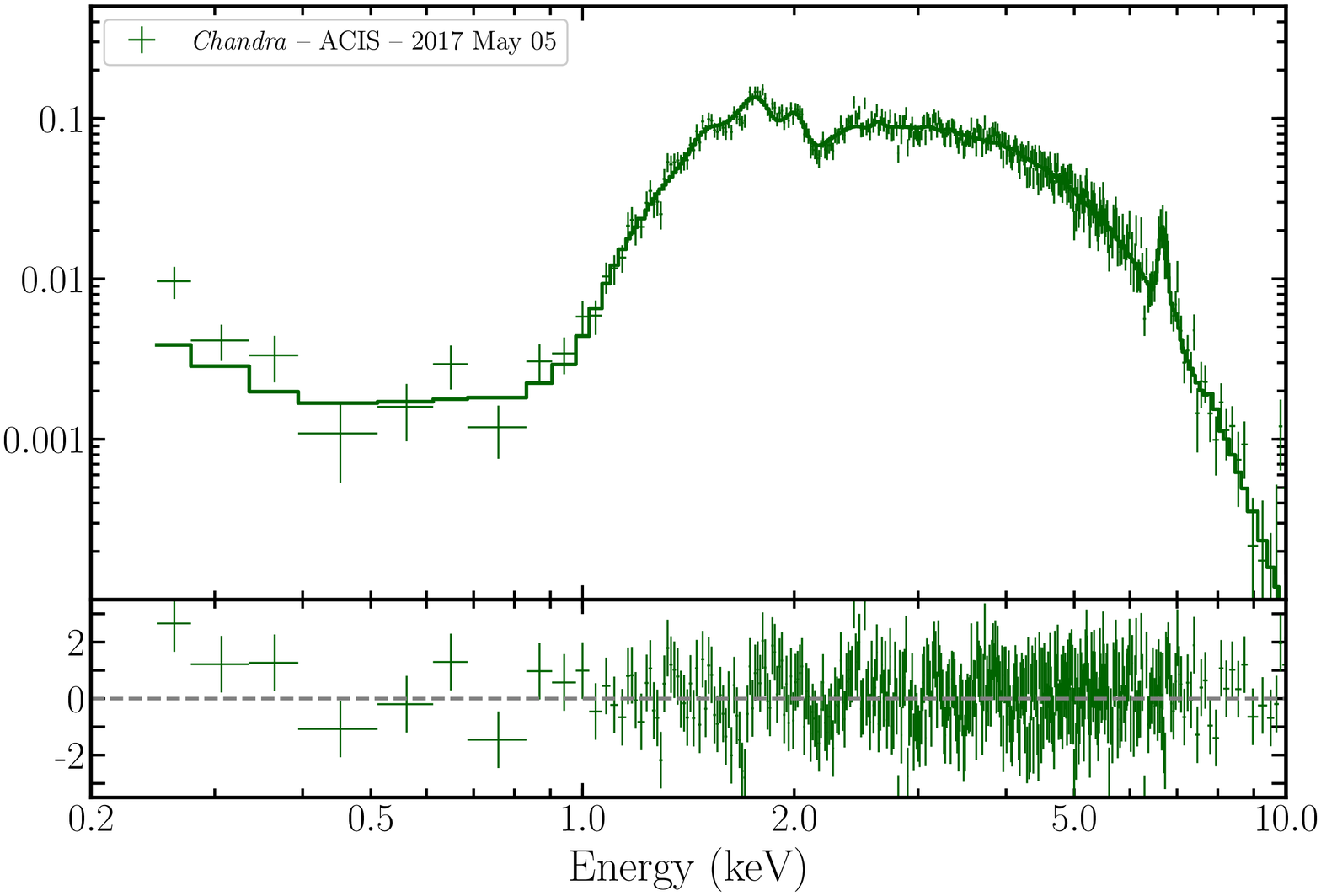} \\
\end{array}$
 \caption{The folded X-ray spectra observed by \emph{XMM-Newton} and \emph{Chandra} with the best fit thermal plasma $\tt{apec}$ model over-plotted, using the parameters listed in Supplementary Information Table\,\ref{tab:x_ray_fits}. The legend in each panel reports the observing X-ray telescope, X-ray instrument, and the epoch of the observation. The $\chi$-values for the model fits are displayed below the spectrum for each epoch, and are coloured corresponding to the X-ray instrument. pn, MOS1, MOS2, and ACIS are coloured black, red, blue, and green, respectively. The uncertainties in the plots correspond to 1-$\sigma$.}
\label{fig:xray_seds}
\end{center}
\end{figure*}

\begin{table*}
\renewcommand\tablename{Supplementary Information Table}
  \small
  \caption{\label{tab:x_ray_fits} The best-fit parameters of the $\tt{apec}$ model to the X-ray observations of Apep. All reported errors are for 90\% confidence. The abundances are relative to solar abundances \citep{Wilms2000}. $S_{\rm{x},\rm{obs}}$ and $S_{\rm{x},\rm{unabs}}$ are the observed and unabsorbed 0.2 to 10.0\,keV flux, respectively.}
  \begin{center}
    \begin{tabular}{cccccc}
    \hline
Epoch & $kT$ & $N_{\rm{H}}$ & Abundances & $S_{\rm{x},\rm{obs}}$ $S_{\rm{x},\rm{unabs}}$ & $\chi^{2}$/dof ($\chi_{\rm{red}}^{2}$) \\
 & (keV) & (10$^{22}$\,cm$^{-§2}$) & & (10$^{-12}$\,ergs\,cm$^{-2}$\,s$^{-1}$) & \\
    \hline    
    \hline    
2004 Aug 10 & 5.6\,$\pm$\,0.5 & 2.8\,$\pm$\,0.1 & 0.5\,$\pm$\,0.1 & 9.2\,$\pm$\,0.2~~17.8\,$\pm$\,0.6 & 551/510 (1.08)\\
2006 May 21 & 4.6\,$\pm$\,0.3 & 3.10\,$\pm$\,0.09 & 0.7\,$\pm$\,0.1 & 8.4\,$\pm$\,0.2~~17.1\,$\pm$\,0.4 & 392/352 (1.11)\\
2008 Mar 20 & 6.3\,$\pm$\,0.4 & 2.7\,$\pm$\,0.1 & 0.8\,$\pm$\,0.1 & 9.6\,$\pm$\,0.2~~17.6\,$\pm$\,0.3 & 821/677 (1.21)\\
2015 Mar 08 & 5.1\,$\pm$\,0.2 & 2.76\,$\pm$\,0.05 & 0.48\,$\pm$\,0.04 & 8.1\,$\pm$\,0.1~~16.3\,$\pm$\,0.2 & 1214/858 (1.42)\\
2017 May 02 & 4.6\,$\pm$\,0.3 & 2.86\,$\pm$\,0.08 & 0.48\,$\pm$\,0.04 & 8.2\,$\pm$\,0.2~~16.2\,$\pm$\,0.2 & 423/388 (1.09)\\
2017 May 05 & 4.7\,$\pm$\,0.3 & 2.89\,$\pm$\,0.08 & 0.53\,$\pm$\,0.08 & 8.1\,$\pm$\,0.2~~16.2\,$\pm$\,0.2 & 464/390 (1.19)\\

    \hline\end{tabular}
\end{center}
\end{table*}

The Fe\,K$\alpha$ line profile does not change significantly between the observations, occurring at 6.69\,$\pm$\,0.01\,keV. There are also statistically significant lines at 1.96\,$\pm$\,0.04 and 2.44\,$\pm$\,0.03\,keV corresponding to Si\,K$\alpha$ and S\,K$\alpha$, respectively.

As is evident by the different values of the observed 0.2 to 10\,keV flux, $S_{\rm{x},\rm{obs}}$, Apep shows low long term ($>1.5$\,year) X-ray variability. Such low long term variability, combined with the bright X-ray luminosity, is consistent with temperature of the collisional plasma changing in a CWB system with a period $\gtrsim 6$ months \citep{2011ApJ...727L..17Z}. A search for variability internal to the \xmm~observations was also conducted. We performed a $\chi^{2}$ and Rayleigh test\citep{Callingham2012} to identify any variability or a period signal within the range of 6.4\,s to 12\,h for the MOS1/MOS2 data, and 146\,ms and 12\,h for the pn data. We detected no significant variability and saw no significant power at any period. Additionally, the 2004 and 2008 \xmm~observations did not show time variability on timescales greater than 10\,s according to a $\chi^{2}$ test performed by the \xmm~pipeline \citep{Watson2009}. The lack of short-term variability makes the presence of a compact object in the system unlikely \citep{2011ApJ...727L..17Z}. Finally, we note that in a study of 4330 of the brightest, compact sources in the 2XMMi-DR3 catalog \cite{Watson2009}, Apep was identified as a compact object system \cite{2012ApJ...756...27L}. The authors arrived at such a classification, over a CWB, because the authors did not consider multi-wavelength properties of Apep, particularly its infrared properties, comprehensively model the X-ray spectrum, and were heavily influenced by the low Galactic latitude of Apep.

\subsection{Radio properties of Apep}

From the ATCA measurements of Apep, the flux density $S$ was observed to increase from $27.9 \pm 0.9$\,mJy at frequency $\nu$ of 19.7\,GHz to $166 \pm 15$\,mJy at 1.4\,GHz. The radio spectrum is well described by a power-law with a spectral index of $\alpha = -0.71 \pm 0.05$, where $S \propto \nu^{\alpha}$. Apep was unresolved in all of the radio observations, with the highest resolution observation at 19.7\,GHz having a synthesised beam profile of $0.74'' \times 0.29''$. 

Such flux densities makes Apep the brightest non-thermal radio CWB discovered by over an order of magnitude, and the second brightest CWB detected in the radio outside of $\eta$~Carinae \citep{2013A&A...558A..28D}. The relatively faint ($< 20$\,mJy) radio flux densities of CWBs is usually explained by a combination of free-free absorption, lack of strong magnetic fields, and relatively weak shocks \citep{2002ApJ...566..399M,2013A&A...558A..28D}. For Apep to be so bright in the radio, it is possible free-free absorption is significantly less than normal, which is consistent with asymmetric mass-loss model since the majority of potentially obscuring medium would be in a plane rotated to our line of sight to the shock region. Alternatively, a significantly stronger magnetic field ($\sim$\,20\,kG) could be present in the system. Such a strong magnetic field is also predicted by rapid rotator models for WR stars \citep{2014A&A...562A.118S}.

Apep was also observed at 843\,MHz by the Molonglo Observatory Synthesis Telescope (MOST) on six separate occasions between 1988 to 2006, mostly as part of the second epoch Molonglo Galactic Plane Survey-2 (MGPS-2) \citep{Bock1999,Murphy2007}. The flux densities of Apep in these images were calculated from MGPS-2 calibrated images and are presented in Supplementary Information Figure\,\ref{fig:radio_lightcurve}. The lightcurve demonstrates that the Apep is variable at 843\,MHz, with the flux density of Apep increasing from 85\,mJy to 138\,mJy between 1988 and 2006. The increase in the flux density appears to be linear with time, lending further evidence of a long-term secular evolution of the emission from Apep, similar in time scale as implied by the wrapping of the dust spiral in the VISIR data. The type of linear variability displayed in Supplementary Information Figure\,\ref{fig:radio_lightcurve} also rules out refractive interstellar scintillation as the cause \citep{Rickett1984}. The significant long term radio variability seems incongruent with the low X-ray variability. It is possible that the significant radio variability could be due to variation of the free-free absorption medium over the orbit \citep{2015A&A...579A..99B}, consistent with the model proposed in the main text of a greater than decade period orbit that takes the colliding-wind companion of the Central Engine through the dense equatorial plane populated by the slow, cool wind of the rapid rotator.

\begin{figure*}[h]
\renewcommand\figurename{Supplementary Information Figure}
\begin{center}
    \includegraphics[scale=0.285]{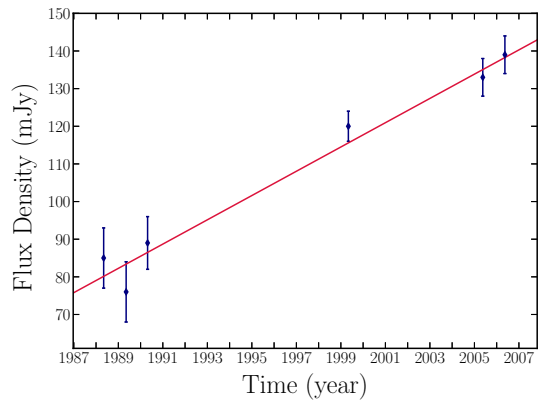}
    \caption{843\,MHz radio lightcurve of Apep between 1988 and 2006. The red line is the best fit to the data, and predicts a flux density of $175\pm15$ mJy for May 2017, when the targeted ATCA observations were performed. The gradient of the line implies the source brightens 3.2\,mJy\,yr$^{-1}$ at 843\,MHz. The uncertainties represent 3-$\sigma$.}
    \label{fig:radio_lightcurve}
\end{center}
\end{figure*}

\end{document}